\documentclass[twocolumn]{aastex631}

\usepackage{booktabs}
\usepackage{array}
\usepackage{multirow}
\usepackage{wrapfig}
\usepackage{threeparttable}
\usepackage{graphicx}
\usepackage{xcolor}
\usepackage{hyperref}
\usepackage{subcaption}
\usepackage{amsmath}
\usepackage{newunicodechar}
\makeatletter
\let\tablenum\@undefined
\makeatother
\usepackage[detect-all]{siunitx}

\usepackage{xspace}
\hypersetup{linkcolor=cyan,citecolor=blue,filecolor=cyan,urlcolor=magenta}

\newcommand{\as}{$^{\prime\prime}$\xspace}

\makeatletter
\@ifundefined{longtable*}{}{%
  \expandafter\let\csname longtable*\endcsname\@undefined
  \expandafter\let\csname endlongtable*\endcsname\@undefined
}
\makeatother

\begin{document}

\title{The magnetic field in the multi-phase interstellar medium of NGC\,6946}

\author[0009-0002-5505-9985]{Amir Yarahmadi}
\affiliation{School of Astronomy, Institute for Research in Fundamental Sciences (IPM), P.O. Box 1956836613, Tehran, Iran}
\email{amir.yarahmadi73@gmail.com}

\author[0000-0002-0377-0970]{Fatemeh Tabatabaei}
\affiliation{School of Astronomy, Institute for Research in Fundamental Sciences (IPM), P.O. Box 1956836613, Tehran, Iran}
\affiliation{I. Physikalisches Institut, University of Cologne, 50937 Cologne, Germany}
\email{fatemeht@gmail.com}

\author{Golshan Ejlali}
\affiliation{School of Astronomy, Institute for Research in Fundamental Sciences (IPM), P.O. Box 1956836613, Tehran, Iran}
\email{glshnejlali@gmail.com}

\author{Rainer Beck}
\affiliation{Max-Planck-Institut f\"ur Radioastronomie, Auf dem H\"ugel 69, 53121 Bonn, Germany}
\email{rbeck@mpifr-bonn.mpg.de}

\author[0000-0001-5357-6538]{Enrique Lopez-Rodriguez}
\affiliation{Department of Physics \& Astronomy, University of South Carolina, Columbia, SC 29208, USA}
\email{lopezroe@mailbox.sc.edu}

\author[0000-0003-3249-4431]{Alejandro Borlaff}
\affiliation{NASA Ames Research Center, Moffett Field, CA 94035, USA}
\affiliation{Bay Area Environmental Research Institute, Moffett Field, California 94035, USA}
\affiliation{IAU Centre for the Protection of the Dark and Quiet Sky, 98-bis Blvd Arago, 75014 Paris, France}
\email{a.s.borlaff@nasa.gov}

\begin{abstract}
Magnetic fields regulate the multi-phase interstellar medium (ISM) of spiral galaxies, yet their behavior across ISM phases remains poorly constrained
observationally. We present a comparative polarimetric study of the nearby spiral galaxy NGC\,6946 using far-infrared (FIR) data at 154\,$\mu$m from SOFIA/HAWC+ and radio
continuum data at 6\,cm from the VLA and Effelsberg telescopes, which probe magnetic fields in the cold dust-emitting and diffuse synchrotron-emitting phases of the ISM,
respectively.

The synchrotron polarization reveals a coherent large-scale spiral magnetic field, with high polarization fractions in interarm regions and strong depolarization in star-forming
arms. The FIR polarization fraction is lower on average and shows stronger spatial variations. Both tracers decrease systematically with increasing star formation
activity, indicating that star-formation-driven turbulence reduces magnetic field ordering; the synchrotron polarization fraction is strongly suppressed in the central kiloparsec but stable across the mid disk, with a tendency to increase beyond $R \approx 7$\,kpc, while the FIR polarization fraction remains low ($\sim$2--3\%) and approximately constant over the radial range where it can be reliably measured ($R \lesssim 5.4$\,kpc).

Our results indicate that NGC\,6946 hosts a globally ordered magnetic field locally modulated by turbulence. This study highlights the complementary nature of
FIR and radio polarimetry in probing magnetic fields across ISM phases and provides quantitative constraints on how magnetic field ordering varies between the cold and diffuse ISM
components.
\end{abstract}


\section{Introduction}\label{sec:intro}

Magnetic fields are fundamental constituents of the interstellar medium (ISM) and are expected to play a major role in the evolution of spiral galaxies.
Magnetohydrodynamic (MHD) theories, in particular mean-field dynamo models, predict that large-scale coherent magnetic fields can be amplified and maintained in spiral galaxies through the combined effects of differential rotation and turbulence \citep{Beck2015, Subramanian1998, Brandenburg}. They are also thought to affect star formation by providing magnetic pressure, controlling cloud fragmentation
\citep{Tabatabaei2018, Pattle2022}, and mediating stellar feedback. In particular, theoretical studies suggest that magnetic fields can shape large-scale spiral structure \citep{Kim_2002} and
regulate star formation activity \citep{mac2004}, potentially playing an important role even in the earliest phases of galaxy evolution \citep{silk2006}. Constraining these
theoretical predictions requires multi-wavelength observations capable of probing magnetic fields across different phases of the ISM.

Our current knowledge of extragalactic magnetic fields is largely based on a combination of radio continuum and far-infrared (FIR) polarimetry. At centimeter wavelengths, synchrotron emission from cosmic-ray electrons spiralling around magnetic field lines is intrinsically linearly polarized and provides a tracer of the plane-of-sky magnetic field orientation in the diffuse interstellar medium. The total synchrotron intensity constrains the strength of the total magnetic field (under the equipartition assumption), while the polarization fraction probes the level of magnetic field ordering. In addition, Faraday rotation measurements allow determination of the line-of-sight component of the regular magnetic field.

In contrast, FIR polarization arises from thermal emission of non-spherical dust grains aligned with the magnetic field, and thus traces the magnetic field structure within the cold and dense phases of the ISM. Dust grain alignment is commonly explained by Radiative Alignment Torque (RAT) theory, in which anisotropic radiation fields align dust grains with respect to the local magnetic field \citep{Andersson2015}. FIR polarimetry therefore provides a complementary view of magnetic fields in regions associated with molecular gas and star formation, where synchrotron emission may be weak or affected by depolarization.

Early attempts to trace extragalactic magnetic fields using dust polarization were carried out at near-infrared wavelengths. \cite{Jones2000} demonstrated that NIR
polarization can trace large-scale magnetic structures in external galaxies. A major advance came with the launch of Planck, which delivered all-sky maps of dust polarization
in the Milky Way and revealed strong correlations between dust and synchrotron polarization \citep{Planck2015}. Although Planck lacked the angular resolution to resolve most
external galaxies in detail, it demonstrated the power of dust polarization as a magnetic field tracer. High-resolution extragalactic FIR polarimetry became possible with the
Stratospheric Observatory for Infrared Astronomy (SOFIA) and its HAWC+ polarimeter. The Survey on extragALactic magnetiSm with SOFIA (SALSA) has provided resolved FIR
polarization maps of several nearby spiral galaxies (\cite{SALSAI}; \cite{SALSAII}). These studies revealed that in some galaxies, such as M51, dust and
synchrotron polarization show a remarkable correspondence, whereas in others, including NGC~6946, the relationship appears more complex, likely reflecting variations in ISM
conditions, turbulence, and the coupling between magnetic fields and different gas phases.

Recent observational studies combining FIR and radio polarimetry have shown that magnetic field properties vary systematically across ISM phases \citep{SALSAI, SALSAV, lopezrodriguez2023, SALSAII}: radio observations typically reveal large-scale fields with high polarization fractions in interarm regions and reduced ordering in star-forming arms \citep{Beck2015}, while FIR polarization often exhibits lower polarization fractions and stronger local variations, reflecting the influence of turbulence, gas density, and star formation activity \citep{SALSAV, Surgent2023}.

\begin{table}[ht]
    \centering
    \begin{threeparttable}
    \caption{General parameters adopted for NGC\,6946.}
    \begin{tabular}{ll}
        \hline
        \textbf{Parameter} & \textbf{Value} \\
        \hline
        Position of nucleus (J2000) & RA = $20^{\mathrm{h}}34^{\mathrm{m}}52.3^{\mathrm{s}}$ \\
                                   & Dec = $60^\circ09'14''$ \\
        Position angle of major axis\tnote{1} & $242^\circ$ \\
        Inclination\tnote{1} & $38^\circ$ ($0^\circ$ = face on) \\
        Distance\tnote{2} & 7.72\,Mpc\\
        \hline
    \end{tabular}
    \label{tab:galaxy-properties}
    \begin{tablenotes}
        \item[1] \cite{Boomsma+2008}
        \item[2] \cite{Eldridge+2019}
        \item[3] $1'$ = 2.25\,kpc along the major axis.
    \end{tablenotes}
    \end{threeparttable}
\end{table}

Among nearby systems, NGC~6946 has become a benchmark object for studying galactic magnetism. Previous radio studies revealed strong total magnetic fields and prominent
ordered components, including well-defined magnetic arms located between the optical spiral arms \citep{Beck+1996}. NGC~6946 is a nearby grand-design Scd spiral galaxy
(Table~\ref{tab:galaxy-properties}) located at a distance of 7.72\,Mpc  \citep{Eldridge+2019}. It is characterized by a high global star formation rate of $\sim 7.1~M_\odot~\mathrm{yr}^{-1}$
\citep{Kennicutt_2011}, numerous luminous H\,{\sc ii} regions \citep{Beck+1996}, and a complex spiral structure with a particularly bright northeastern arm \citep{Beck2007}.
The galaxy shows no compelling evidence for dominant AGN activity (e.g., \citealt{Tasi+2006}), making it well suited for studying the interplay between magnetic fields and star
formation in a star-forming disk environment. These properties make NGC~6946 an excellent laboratory for testing theories of magnetic field amplification and ordering.

Previous multi-wavelength studies of NGC~6946 have shown unique, phase-dependent magnetic field signatures: a highly ordered, large-scale magnetic field traced by radio synchrotron
polarization, together with a highly turbulent magnetic field traced by FIR dust polarization \citep{Beck1991, Beck2007, SALSAV, SALSAII}. However,
these two magnetic field tracers have not previously been analyzed jointly in this galaxy. Here, we perform a detailed multi-phase magnetic field analysis of NGC~6946, combining the FIR and
radio polarimetry to directly compare the two tracers. Our approach includes studying the spatial distribution of pure synchrotron polarization fraction (p) and
quantifying differences in magnetic field ordering between the cold dust-emitting and the synchrotron-emitting media, providing new observational constraints on this coupling. We also investigate dependencies on star formation rate, gas density,
turbulence, and radiation field.

This paper is organized as follows. In Section~2, we describe the observational data sets and the preprocessing steps applied to the FIR and radio maps. Section~3 presents
the overall distributions of total and polarized emission. Section~4 examines the radial behavior of the polarized intensity, while Section~5 introduces the analysis of the
polarization fraction in both synchrotron and dust emission, including their radial variations. Section~6 discusses the implications of our results in the context of
magnetic field amplification, ISM structure, star formation, and provides a quantitative comparison between dust and synchrotron polarization properties. Finally, Section~7
summarizes our main conclusions. The separation of the thermal and nonthermal radio components at 6\,cm, which underpins the synchrotron polarization analysis of Section~5,
is described in Appendix~\ref{TNT}.

\section{Data} \label{sec:data}

\begin{figure*}[ht!]
\centering
\includegraphics[width=0.48\textwidth]{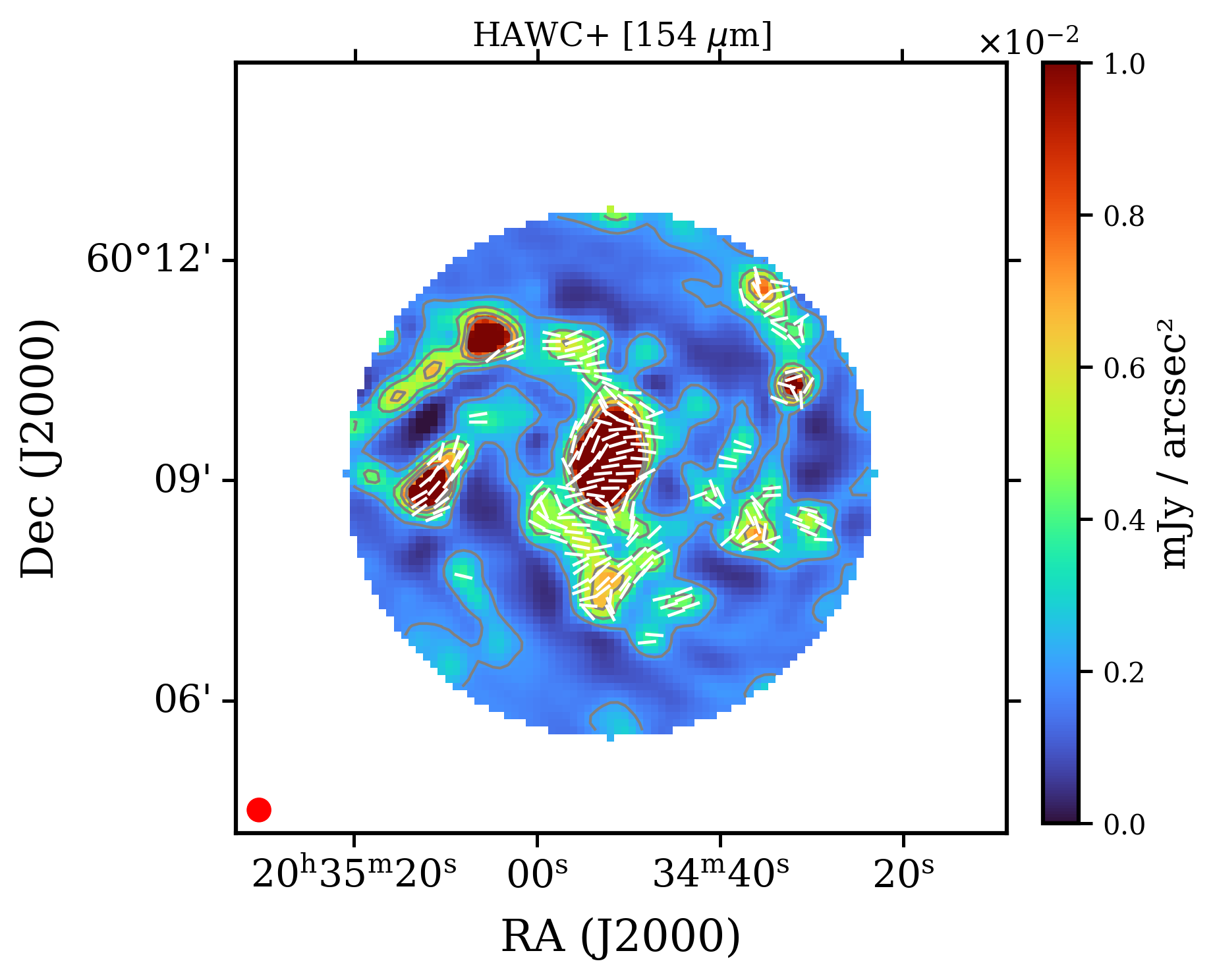}
\hfill
\includegraphics[width=0.48\textwidth]{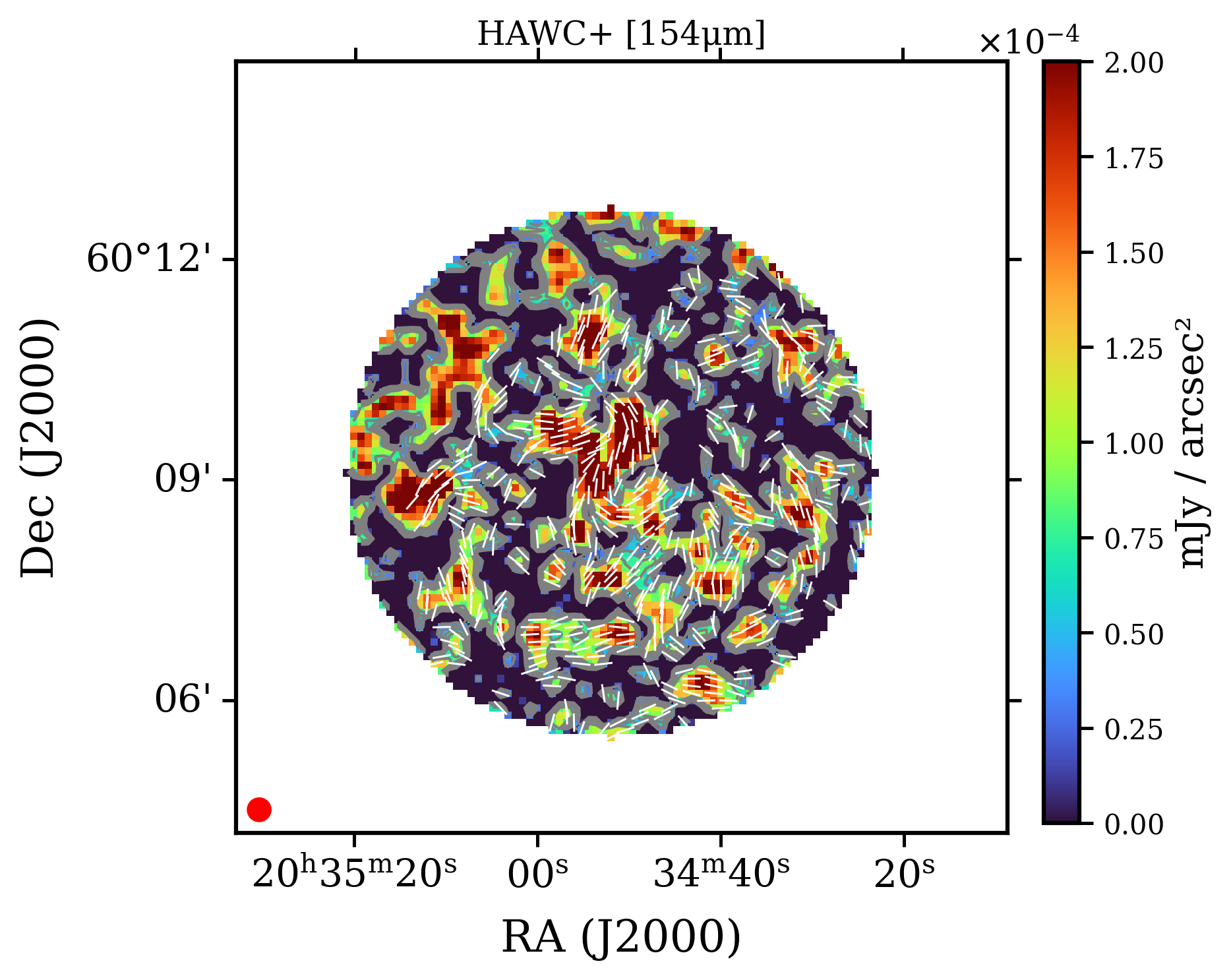}
\includegraphics[width=0.48\textwidth]{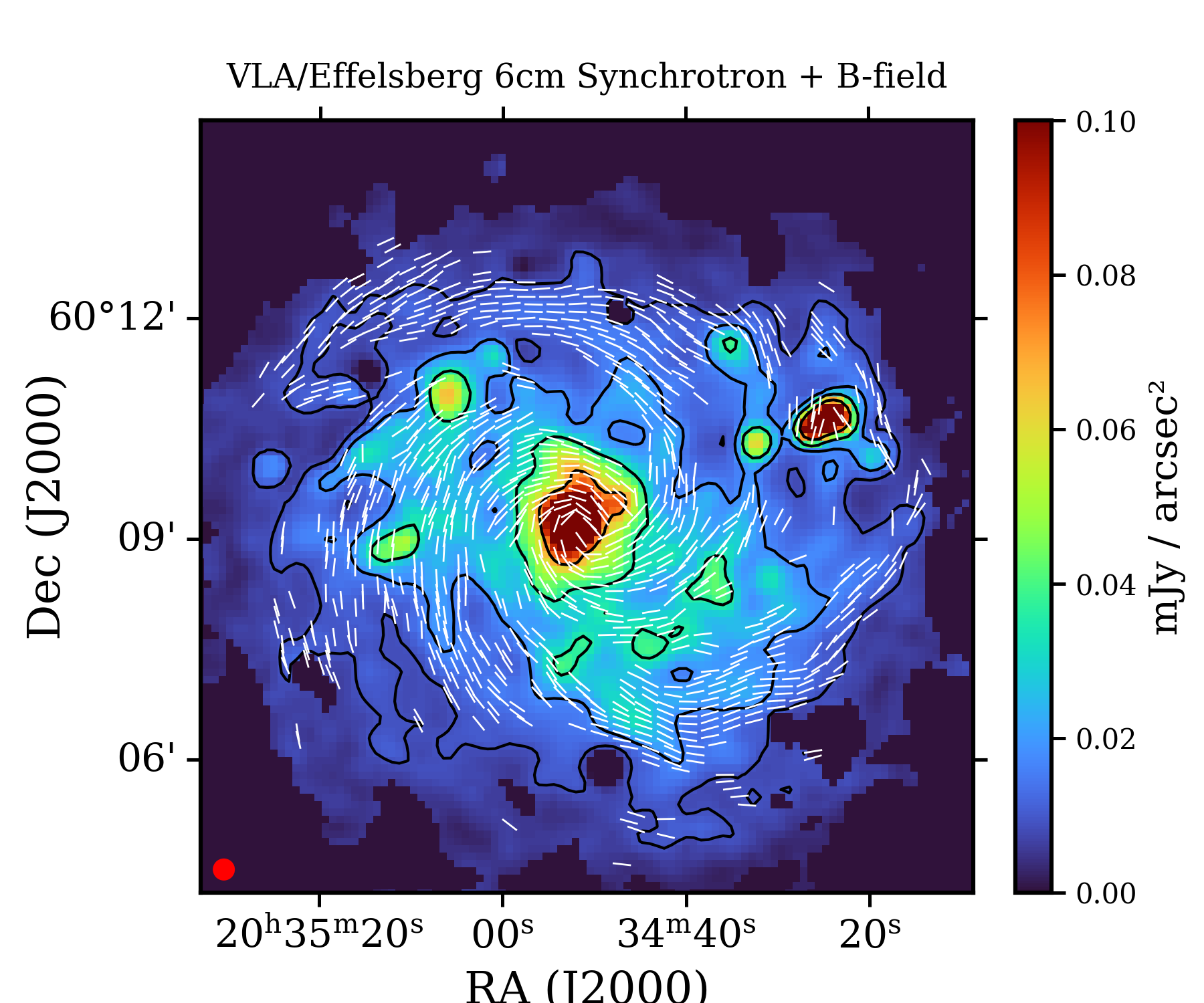}
\hfill
\includegraphics[width=0.48\textwidth]{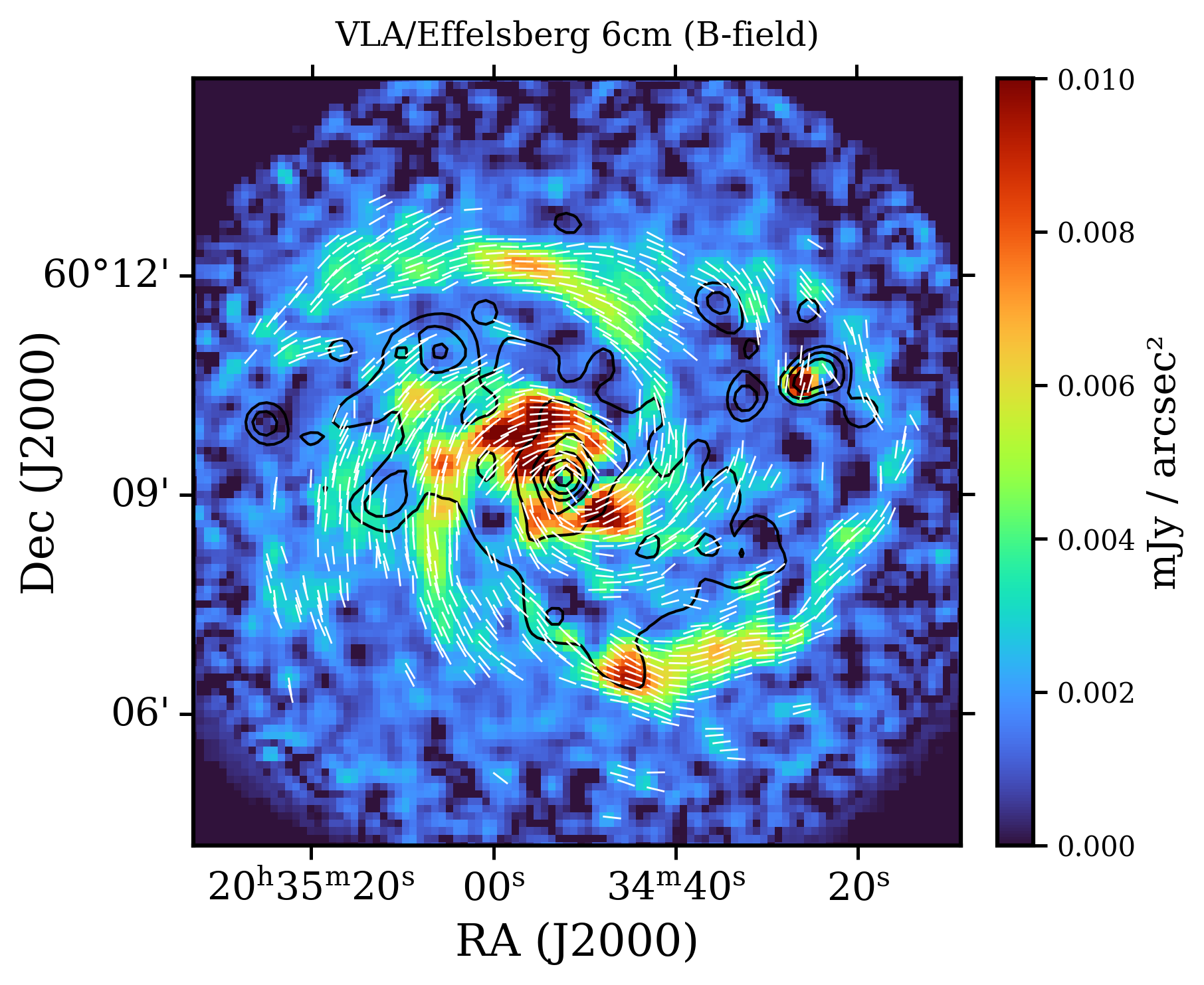}
\caption{Top: {\it SOFIA/HAWC+} dust polarization
orientations (at  154\,$\mu$m) overlaid on top of the  dust total intensity ($I_{\rm dust}$, left) and polarized intensity ($PI_{\rm dust}$, right) maps. Contours on the left
show the 1.5, 3, 6, 12$\times \sigma_{\rm rms}$ levels of the $I_{\rm dust}$ map, and on the right 1.5, 3, 6, 12 $\times \sigma_{\rm rms}$ levels of the $PI_{\rm dust}$ map.
The noisy areas at the edges of the maps were clipped.
Bottom: Radio synchrotron (6\,cm) polarization
orientations overlaid on top of the synchrotron total intensity ($I_{\rm sync}$, left) and polarized intensity ($PI_{\rm sync}$, right) maps. Contours on the left show 16,
32, 64, 128 $\times \sigma_{\rm rms}$ levels of the $I_{\rm sync}$ map, and on the right 16, 32, 64, 128 $\times \sigma_{\rm rms}$ levels of the $PI_{\rm sync}$ map. Circles
on the bottom left corners indicate the angular resolutions of 18$\arcsec$.}

\label{fig:ngc6946_maps}
\end{figure*}

The data used in this work are explained as follows and summarized in Table \ref{tab:NGC~6946_data}.
NGC\,6946 was observed in FIR and radio regimes with SOFIA and VLA telescopes in various wavelengths. In Figure~\ref{fig:ngc6946_maps} we present a detailed comparison
between the radio continuum emission at 6\,cm (4.85\,GHz) and the FIR emission at 154\,\micron\ for the face-on spiral galaxy NGC\,6946.

\begin{table*}
    \centering
    \caption{{Images of NGC\,6946 used in this study.}}
    \begin{tabular}{lllll}
        \toprule
        Image & rms noise & Resolution & Telescope & Ref. \\
        \midrule
         Radio-6~cm  &  & 15$''$ & VLA + Effelsberg&\cite{Beck+1996}\\
           Stokes-I &22$\mu$Jy/beam & &        &  \\
           Stokes-Q &15$\mu$Jy/beam & &       &  \\
           Stokes-U & 15$\mu$Jy/beam & &       &  \\
           \hline
         FIR-154$\mu$m  &    & 13.6$''$ & SOFIA/HAWC+ &\cite{SALSAIII}\\
        Stokes-I &0.04~Jy/pixel  &    &   & \\
       Stokes-Q & 0.02~Jy/pixel& &      &  \\
       Stokes-U & 0.02~Jy/pixel & &       &  \\
       \hline
       HI-21\,cm  & 1.4 Jy/beam\,m\,s$^{-1}$  &6$''$   & VLA & \cite{walter+2008} \\
        CO(2–1) &  0.06\,K\,km\,s$^{-1}$  &13 $''$ & IRAM 30-m  & \cite{Leroy+2009} \\
        $U_{\rm min}$  & 0.5 & 18\as & Herschel\& Spitzer& \cite{Aniano2012} \\
        H$\alpha$ (de-reddened)  & 0.06\,$\mu$~erg~s$^{-1}$cm$^{-2}$ sr$^{-1}$ & 18\as & KPNO & \cite{Tabatabaei+2013} \\
        SFR  & 0.004\,M$_{\odot}$\,kpc$^{-2}$\,yr$^{-1}$ & 12''  &GALEX \& WISE-W4 & \cite{Salim2016} \& \cite{Leroy2019} \\
        \bottomrule
    \end{tabular}%
    \label{tab:NGC~6946_data}
\end{table*}


\subsection{SOFIA HAWC+ data}

NGC\,6946 was observed in the FIR regime with \textit{SOFIA} as part of the \textit{Survey on extragALactic magnetiSm with SOFIA} (SALSA Legacy Program), as
detailed in \cite{SALSAIII}. The observations were carried out at 154\,$\mu$m (D-band) using the on-the-fly mapping (OTFMAP) technique in polarimetric mode with
\textit{HAWC+}. This method, which traces a Lissajous scan pattern to keep the target within the field-of-view at all times, significantly improves observational efficiency
and sensitivity compared to traditional chopping and nodding modes. This dataset has been previously presented and analyzed as part of the SALSA survey (e.g., \citealt{SALSAIV, SALSAV}).

The raw data were reduced using the \textit{Comprehensive Reduction Utility for SHARP II} (CRUSH; \citealt{Kovacs+2008}) version 2.50-1, along with the HAWC+ Data Reduction
Pipeline (DRP) version 2.7.0, developed by the Data Reduction Pipeline Group at the \textit{SOFIA} Science Center. The data reduction follows the procedures described in \cite{SALSAIII}, incorporating the latest calibration and producing homogeneously reduced polarimetric maps suitable for scientific analysis.

The HAWC+ pipeline provides uncertainty maps associated with each Stokes parameter for every pixel. These error maps were used directly in our analysis to quantify the statistical uncertainties in the Stokes $Q$ and $U$ maps. The polarized intensity was corrected for positive bias using the standard debiasing method,
\begin{equation}
P = \sqrt{Q^2 + U^2 - \sigma_P^2},
\end{equation}
where $\sigma_P$ represents the uncertainty in the polarized intensity.
The HAWC+ maps (Fig.~\ref{fig:ngc6946_maps}, top row) are expressed in units of Jy/pixel and have an original angular resolution of 13.6\as and a pixel size of
6.9\as.
%
We estimated the statistical uncertainty of the Q and U maps by measuring the $\sigma_{\rm rms}$ noise in regions of the map far from the galaxy signal.
{The rms noise of Stokes IQU maps at 154\,$\mu$m are listed in Table~\ref{tab:NGC~6946_data}.} We then computed the debiased polarized intensity using the relation provided
in \cite{SALSAI}.

\subsection{Radio data}

We used the 6\,cm radio polarimetric maps with an angular resolution of
15$^{\prime\prime}$ and a pixel size of 5$^{\prime\prime}$, as presented by \citet{Beck+1996}. These datasets were derived from combined observations conducted with the Very
Large Array (VLA) and the Effelsberg 100\,m single-dish radio telescope. Longer wavelengths, specifically 20\,cm, were excluded from our analysis due to the significant
impact of Faraday rotation \citep{Beck1991}, which can distort polarization measurements at these wavelengths. The rms noise of Stokes IQU maps at 6\,cm are listed in
Table~\ref{tab:NGC~6946_data}.
For our analysis, Stokes IQU maps were convolved from their native resolution to 18$^{\prime\prime}$ resolution and resampled to a common pixel size of 6$^{\prime\prime}$
($\sim$ 224~pc). The maps of total and  polarized intensity (see Sect.~\ref{general}) maps at 6\,cm are shown in Fig.~\ref{fig:ngc6946_maps} (bottom row).
%

\subsection{CO and HI data}

To investigate possible connections between neutral gas and polarized intensity (and polarization fraction) in the FIR and radio regimes, 
magnetic field structure,
we use the CO(2--1)  data of NGC\,6946 from the HERACLES survey \cite[HERA CO Line Extragalactic Survey;][]{Leroy+2009}. These data are also part of \textit{The HI Nearby
Galaxy Survey} (THINGS) and the \textit{Spitzer Infrared Nearby Galaxies Survey} (SINGS), providing complementary information on the molecular gas distribution. We used
moments
0 (integrated emission line) and 2 (intensity weighted velocity dispersion) of the CO(2--1) emission line.

The H~I data were obtained from The \textit{H\,\textsc{i} Nearby Galaxy Survey} (THINGS\footnote{\url{https://www2.mpia-hd.mpg.de/THINGS/Data.html}}). As described by
\cite{walter+2008}, these data provide high-resolution 21\,cm spectroscopy of nearby galaxies, including NGC\,6946, which is to trace the distribution and kinematics of the
atomic hydrogen gas. 
our analysis, both moment 0 and 2 maps were convolved with a Gaussian kernel to match the 6\,cm VLA beam size of 18$\arcsec$, and subsequently reprojected onto the same grid
as the VLA 6\,cm observations.

We estimated the total hydrogen column density, $N_{\mathrm{H\,I}} + 2N_{\mathrm{H}_2}$, using integrated emission (moment 0) maps of atomic hydrogen (H\,\textsc{i}) and
molecular gas traced by $^{12}$CO(2--1) for NGC\,6946.

The atomic hydrogen column density, $N_{\mathrm{H\,I}}$, was calculated using the following relation:

\begin{equation}
    N_{\mathrm{H\,I}} = 1.105 \times 10^{21} \,  \left( \frac{I_{\mathrm{H\,I}}}{\mathrm{FWHM}_{\mathrm{HAWC+}}} \right) \, \mathrm[{cm}^{-2}]
\end{equation}
where $I_{\mathrm{H\,I}}$ is the integrated emission line (moment 0) of H\,\textsc{i}, in units of Jy\,beam$^{-1}$\,m\,s$^{-1}$, and $\mathrm{FWHM}_{\mathrm{HAWC+}}$ is the
beam size of HAWC+ at 154\,$\mu$m, in units of arcseconds. Then this map was convolved
to 18$^{\prime\prime}$
resolution.
The molecular hydrogen column density, $N_{\mathrm{H}_2}$, was estimated using the relation:

\begin{equation}\label{eq2}
  N_{2\mathrm{H}_2} = X_{\mathrm{CO}} \, I_{\mathrm{CO}} \, \mathrm[{cm}^{-2}]
\end{equation}
by \cite{Bolatto+2013}, where $I_{\mathrm{CO}}$ is the integrated emission line (moment 0) of CO(2--1), in units of K\,km\,s$^{-1}$, and $X_{\mathrm{CO}}$ is the CO-to-H$_2$
conversion factor, taken to be $2 \times 10^{20}$ cm$^{-2}$\,(K\,km\,s$^{-1}$)$^{-1}$. It is important to note that, since we are using the CO(2--1) intensity map, we first
convert it to CO(1--0) using an appropriate line ratio before applying Equation~\ref{eq2} to derive the \(N_{\mathrm{H}_2}\) map.
 The CO~\textit{J} = 2~$\rightarrow$~1 to \textit{J} = 1~$\rightarrow$~0 line intensity ratio (\(R_{21}\)) typically ranges from 0.6 to 1.0, similar to what is observed in
 the Milky Way and other nearby galaxies. Central regions of galaxies often exhibit higher values. An average ratio of around 0.8 can be explained by molecular gas that is
 optically thick and has an excitation temperature near 10\,K \citep{Leroy+2009}.
  {The final column density is then }
  $N_{\mathrm{H\,I}} + 2N_{\mathrm{H}_2} = N_{\mathrm{H\,I}} + N_{2\mathrm{H}_2}$.

\subsection{Other data}

{To investigate correlations with star formation rate (SFR), we use a two wavelength hybrid tracer to estimate the SFR surface density, $\Sigma_{\rm{SFR}}$, which correct the dust attenuation in the UV wavelength range using IR data. Specifically, we used the far-ultraviolet (FUV) and IR images with the coefficients calibrated using the GALEX-SDSS-WISE Legacy Catalog \citep[GSWLC;][]{Salim2016} as estimated in the z0MGS project \citep{Leroy2019}. The $\Sigma_{\rm{SFR}}$ is estimated as

\begin{equation}
\label{eq:DSFR}
\begin{split}
\Sigma_{\mathrm{SFR}} ={}& 8.85 \times 10^{-2}\, I_{\mathrm{FUV}} \\
&+ 3.02 \times 10^{-3}\, I_{\mathrm{W4}} \\
&\quad \left[ M_{\odot}\,\mathrm{yr}^{-1}\,\mathrm{kpc}^{-2} \right]
\end{split}
\end{equation}

where $I_{\rm{FUV}}$ is the intensity at FUV using \textit{GALEX} in units of MJy sr$^{-1}$, and $I_{\rm{W4}}$ is the intensity at W4 using \textit{WISE} in units of MJy sr$^{-1}$. }

{We also use a map of the radiation field ($U_{\rm min}$) in NGC\,6946, derived through modeling dust emission by \cite{Aniano2012} at an angular resolution of 18$\arcsec$.
This model
assumes that dust grains are illuminated by a mixture of diffuse and intense starlight fields \citep[][see Appendix~\ref{app:umin} for more details]{Draine2007}. This model also resulted in
a dust mass map, using which \cite{Tabatabaei+2013} de-reddened the H$\alpha$ emission in the disk of NGC\,6946 that enables us to separate the thermal and nonthermal radio
continuum emission at 6\,cm in the present study (see Appendix~\ref{TNT}).}

{The FIR and radio maps were smoothed to 18\as
resolution using a Gaussian kernel. All the maps
were normalized to the same grid, geometry, and size before comparison.}




\section{Overall distributions}\label{general}


Figure~\ref{fig:ngc6946_maps} shows the observed SOFIA/HAWC+ map of the total intensity (Stokes-I) at 154\,$\mu$m. The center of the galaxy as well as the star-forming
regions appear as strong sources of the 154\,$\mu$m emission. NGC6946's spiral arms can also be traced in the SOFIA map. In the inter-arm regions, the diffuse emission is
weak $\leq3\sigma$ level due to the limited sensitivity of the SOFIA/HAWC+ observation. The FIR polarized intensity PI shows a patchy, spatially discontinuous
distribution.
The polarization angle PA
fluctuates on small scales (over scales of the resolution).
Most of the star-forming complexes appear as polarized sources, in particular, in the inner disk. The AIPS task \texttt{BLANK} was used to remove non-significant emission
from the 154~$\mu$m image. This procedure masks background noise and low-level artifacts while retaining regions associated with real signal. The blanking was performed
interactively, allowing visual identification of emission structures. Pixels outside these regions were set to blank values, resulting in a cleaner image that highlights the
relevant features for further analysis.

%
Similar to the FIR emission, the radio continuum emission is bright in the center and star-forming regions. The radio map also shows extended diffuse emission at different
intensity levels covering the disk of NGC\,6946 (Fig.~\ref{fig:ngc6946_maps}). Unlike the FIR, the radio polarized emission (rotated by $90^\circ$, representing the magnetic
field orientations) follows a remarkably coherent, large-scale spiral pattern extending across the galactic disk. In addition, it does not show a
correlation with the star-forming regions at 6\,cm.  As discussed by \cite{Beck2007}, the peak of the polarized intensity is often shifted into the inter-arm regions, forming
``magnetic arms'' that trace a well-ordered field distinct from the star-forming gas arms. The synchrotron polarization is strongest in the northern part of the galaxy.

Synchrotron emission follows the optical spiral arms and the central region, indicating enhanced magnetic field strength and cosmic-ray electron density in these zones
\citep{Beck2007, frick2000magnetic}. However, the polarized synchrotron intensity peaks are often displaced into the interarm regions, forming ``magnetic arms'' that
trace a well-ordered field distinct from the star-forming gas arms \citep{Beck2007, moss2013relation}. This shows that while synchrotron emission is sensitive to the
turbulent magnetic field in star-forming regions, its polarized component primarily traces the more uniform, ordered field away from those turbulent zones. The synchrotron
polarization is strongest in the northern part of the galaxy, corresponding to the prominent ``magnetic arm'' \citep{moss2013relation}, and is linked to
the action of a large-scale mean-field dynamo generating an axisymmetric spiral field \citep{chamandy2012galactic}.
 typical strengths of 10--20\,$\mu$G in the arms and 5--10\,$\mu$G in the interarm regions \citep{Beck2007}.

The FIR observations trace the magnetic field geometry within the cold dust disk. The total intensity emission peaks along the spiral arms and central regions, coincident with
star-forming complexes, while the polarized emission appears fainter and more spatially fragmented \citep{SALSAV, martin2023tomographic}. At 154\,$\mu$m,
as mentioned above, the inferred magnetic field does not trace a clear spiral pattern but rather displays complex and locally varying orientations across the disk. To
quantify this, we compute the circular dispersion $\sigma_\psi$ of the FIR-inferred field orientation relative to the local azimuthal (i.e., locally coherent-spiral)
direction, using the same galaxy-plane deprojection and region mask as in Sect.~\ref{sec:gas_phases}. We restrict the analysis to pixels with FIR polarized intensity
detected at ($>3\sigma$). We find $\sigma_\psi = 48\degr$, $76\degr$, and $54\degr$ in the interarm, arm, and center regions, respectively. These large values confirm that the FIR field is only weakly ordered with respect to a coherent large-scale spiral pattern, with the weakest alignment in the arms and
the strongest in the interarm regions.

This morphology suggests that the far-infrared polarization primarily traces the magnetic field within dense clouds and photon-dominated regions, where local feedback,
turbulence, and projection effects can disrupt the large-scale coherence seen in the radio domain.
This fragmented morphology can reflect the influence of turbulence, feedback-driven flows, and differential gas motions within molecular clouds, which tend to randomize
field orientations on sub-kiloparsec scales \citep{SALSAIV}. {We discuss these possibilities in Sect.~\ref{sec:discussion}.}



Because only the nonthermal (synchrotron) component of the 6\,cm emission is linearly polarized, separating it from the thermal free-free emission is essential before the
polarization fraction can be mapped reliably (Sect.~\ref{dp}). We do so following \cite{Tabatabaei+2013}, using a dust-extinction-corrected H$\alpha$ map as a thermal
free-free tracer to obtain the nonthermal (synchrotron) intensity map, $I_{\rm syn}$; the full methodology, the resulting thermal and nonthermal maps, and the regional
thermal fractions are presented in Appendix~\ref{TNT}.

\section{Radial variation of the polarization intensity}
\label{sec:radial_structure}

\begin{figure*}[htb!]
    \centering
    \includegraphics[width=0.75\textwidth]{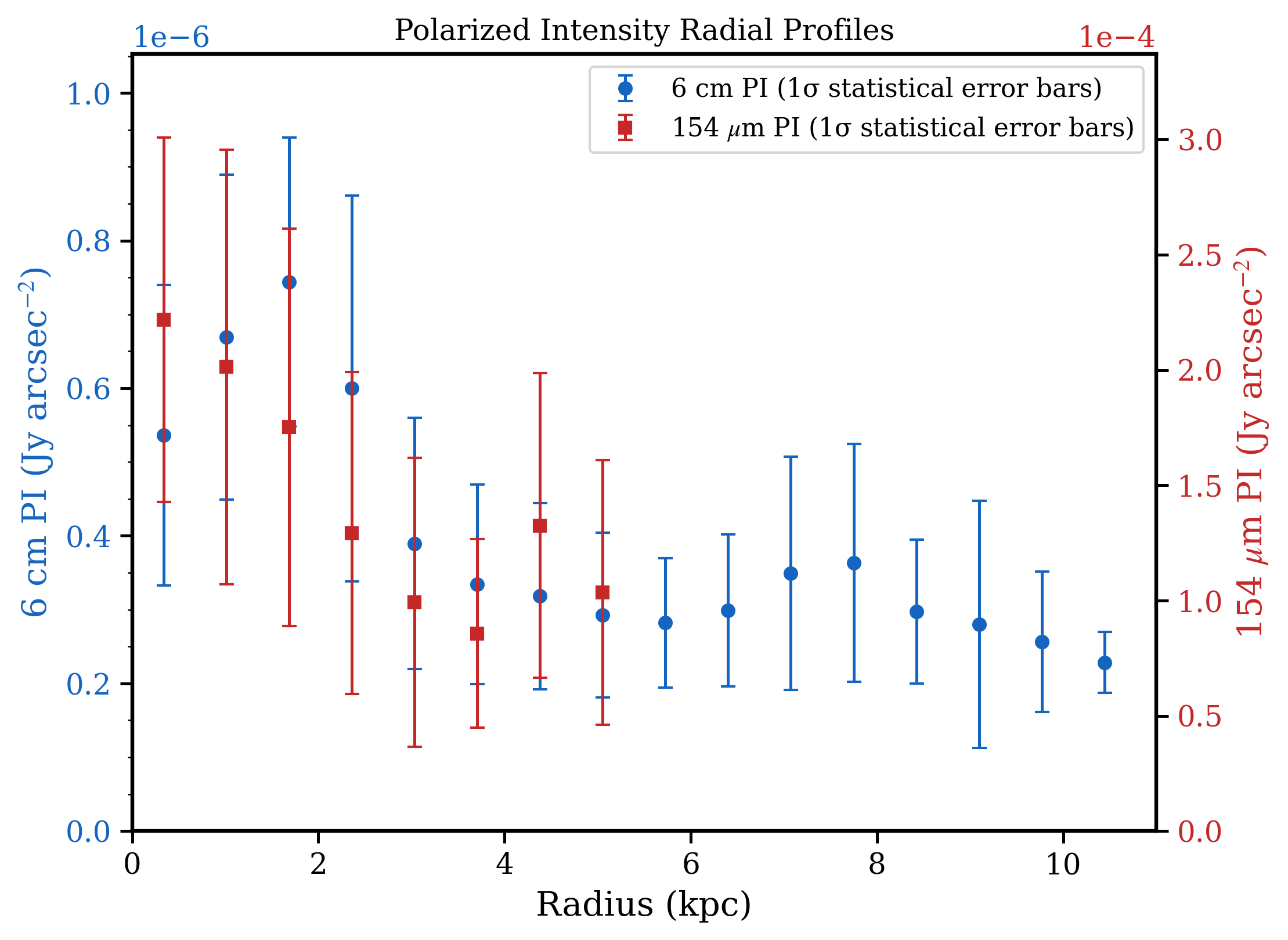}
    \caption{Azimuthally averaged radial profiles of polarized intensity (PI) in NGC\,6946. The 6\,cm radio continuum PI (blue circles, left axis), tracing polarized synchrotron emission,
    shows a broad maximum at $R \sim 1$--$2$\,kpc; the suppressed central value is discussed in Sect.~\ref{sec:radial_structure} in terms of possible Faraday depolarization
    and/or more efficient dust-grain alignment near the nucleus. The 154\,\textmu m far-infrared PI (red squares, right axis), tracing polarized thermal emission from aligned dust grains,
    peaks in the innermost radial bin and declines with radius, with a mild secondary enhancement near $R \sim 4$--$5$\,kpc. Only pixels with total
    intensity detected above $3\sigma_I$ ($\sigma_I = 0.04$\,Jy\,pixel$^{-1}$) are included for the FIR profile, which restricts it to $R \lesssim 5.4$\,kpc --- smaller than both the
    SOFIA/HAWC+ coverage shown in Fig.~\ref{fig:ngc6946_maps} and the radio coverage. Error bars indicate
    the statistical uncertainty (standard deviation within each annulus).}
    \label{fig:pi_profiles}
\end{figure*}

Figure~\ref{fig:pi_profiles} presents the azimuthally averaged radial profiles of the
154~$\mu$m polarized intensity and the 6~cm radio continuum emission,
both convolved to a common angular resolution of $18''$ and expressed
in surface brightness units of Jy\,arcsec$^{-2}$. As with the dust profile below, only pixels with total intensity detected above $8\sigma_I$
($\sigma_I = 22\,\mu$Jy\,beam$^{-1}$; Table~\ref{tab:NGC~6946_data}) are included for the radio profile; in addition, pixels are required to have polarized intensity above $3\sigma_{PI}$
($\sigma_{PI} = \sqrt{2}\,\sigma_{QU} \approx 2.1\times10^{-5}$\,Jy\,beam$^{-1}$, with $\sigma_{QU} = 15\,\mu$Jy\,beam$^{-1}$; Table~\ref{tab:NGC~6946_data}), to avoid
noise-dominated polarized-intensity estimates. The radio profile extends to
$\sim 10$\,kpc. For the dust profile, only pixels with total intensity detected above $3\sigma_I$
($\sigma_I = 0.04$\,Jy\,pixel$^{-1}$; Table~\ref{tab:NGC~6946_data}) are included, since the polarized
intensity is dominated by noise mainly in the outer, low-total-intensity parts of the disk. This restricts the dust profile to
$R \lesssim 5.4$\,kpc, smaller than both the SOFIA/HAWC+ coverage shown in Fig.~\ref{fig:ngc6946_maps}
and the radio coverage.
The radial profiles of PI at 6\,cm and 154\,$\mu$m in NGC\,6946 reveal both common and distinct properties of the magnetized interstellar medium
across the galactic disk. The 6\,cm PI profile shows a broad maximum at $R \sim 1$--$2$\,kpc, followed by a gradual decline toward larger radii and a relatively flat behavior
beyond $\sim 5$\,kpc. This indicates that the synchrotron-emitting cosmic-ray electrons and the ordered large-scale magnetic field are not confined to the inner star-forming
disk, but remain significant out to at least $\sim 10$\,kpc. The weaker PI at radio than FIR at the center of NGC\,6946 may partly reflect Faraday depolarization in
its dense, magnetised interstellar medium affecting the synchrotron polarization. We use ``may'' because the same effect could instead, or in addition, arise if other parameters that favor dust grain alignment
(e.g., the interstellar radiation field) are more effective at the center: if the field there were simply more turbulent, this would suppress the dust polarization as well, so enhanced turbulence alone cannot
fully account for the observed difference. If Faraday depolarization is indeed responsible, the intrinsic synchrotron PI in this region could therefore be larger than observed, and its true radial profile may be more similar
to that at 154\,$\mu$m than the observed profiles suggest. In contrast, the
154\,$\mu$m PI profile is more centrally concentrated, with its highest value in the innermost radial bin, followed by a relatively steep decline out to $R \sim 3$--$4$\,kpc,
after which a mild enhancement appears around $R \sim 4$--$5$\,kpc, near the outer limit of the reliable dust PI measurements. This behavior suggests that the far-infrared polarized emission
is more strongly weighted by the dust column density and the interstellar radiation field, both of which decrease more rapidly with radius than the synchrotron-emitting
cosmic-ray population. Notably, within the statistical uncertainties, the 154\,$\mu$m PI profile also remains broadly constant across the measured radial range, similar to the
6\,cm behaviour beyond the inner few kiloparsecs.

A significant result of this comparison is that both tracers confirm the presence of an ordered magnetic field throughout the disk of NGC\,6946, but they emphasize different
components of the interstellar medium. The 6\,cm PI profile is smoother and more extended, implying that cosmic rays and magnetic fields have a larger radial scale length
than the cool dust responsible for the 154\,$\mu$m polarized emission. The flatter outer radio profile is particularly important, as it suggests that magnetic fields remain
well organized even in regions where the dust emission and star formation activity are already declining. Another notable feature is the dip in the 154\,$\mu$m PI profile
around $R \sim 3$--$4$\,kpc, which may indicate a transition region in the disk where spiral-arm structure, enhanced turbulence, or variations in the field geometry reduce
the net polarized dust emission. The absence of a similarly strong dip in the 6\,cm profile suggests that the large-scale magnetic field traced by synchrotron polarization
remains comparatively coherent on kiloparsec scales, even where the dust polarization is locally reduced.

\section{Polarization fraction variations}\label{dp}

{The polarization fraction is defined as the ratio of the polarized intensity to total power, $p = 100 \times \frac{PI}{I}$.
We obtain maps of $p$ in both radio (synchrotron) and FIR (dust) in NGC\,6946. In the following, we first describe $p_{\rm syn}$, then $p_{\rm dust}$, and finally compare
the differences between the two tracers.}

The polarized emission at 6\,cm is due to only the synchrotron component because the thermal emission does not contribute to $PI$. Studying the variation of $p$ using the
observed radio continuum emission is, thus, misleading neglecting the variation of the thermal fraction over galaxy and considering that it can be significant in spiral arms
and star-forming regions. Therefore, we use the fraction of synchrotron polarization as the physically meaningful parameter to map the fraction of the emission that is
polarized in radio,
$p_{\rm syn} = \frac{PI}{I_{\rm syn}}$.
Valuable information about the ratio between the ordered
and turbulent components of the magnetic field, as well as on depolarization processes along the line
of sight can be obtained through mapping $p_{\rm syn}$ \citep{Tabatabaei2008}.

\begin{figure*}
    \centering
    \begin{subfigure}{0.48\textwidth}
        \centering
        \includegraphics[width=\textwidth]{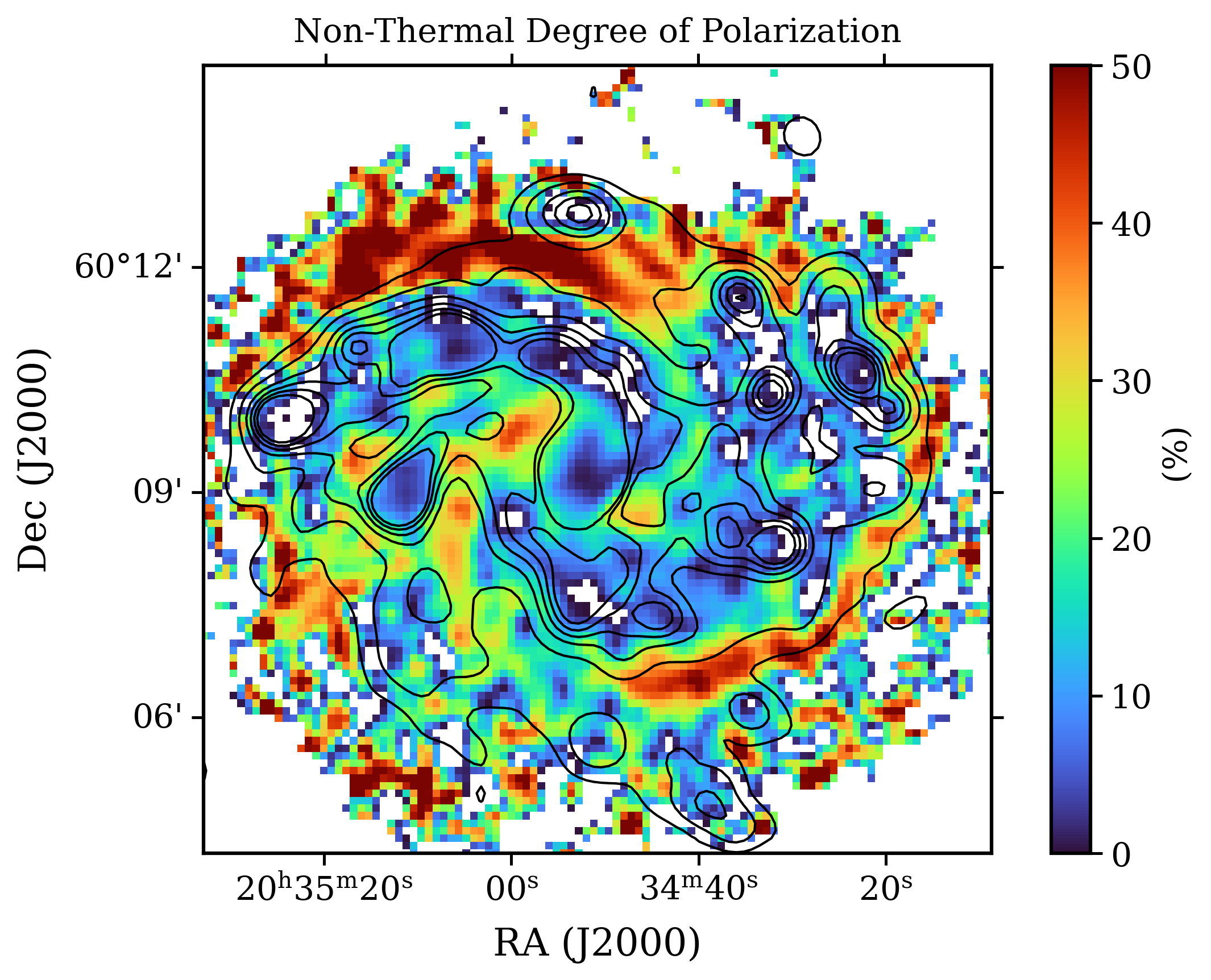}
    \end{subfigure}
    \hfill
    \begin{subfigure}{0.48\textwidth}
        \centering
        \includegraphics[width=\textwidth]{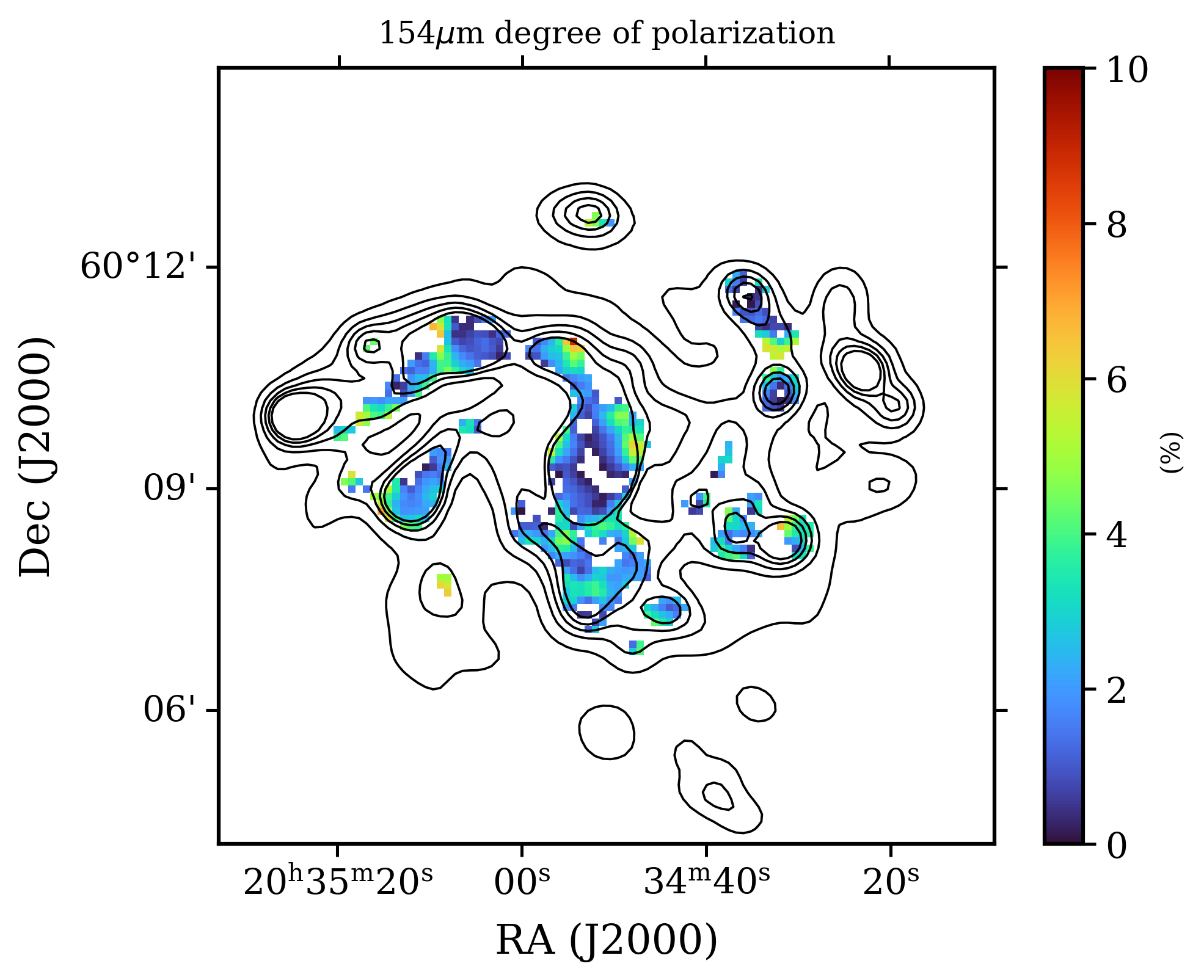}
    \end{subfigure}
    \caption{\textit{Left:} Synchrotron polarization fraction, $p_{\rm syn}$, at $\lambda = 6$\,cm for NGC\,6946. The color bar shows $p_{\rm syn}$ in percentage.
    \textit{Right:} Dust polarization fraction, $p_{\rm dust}$, at $154\,\mu$m for NGC\,6946. The color bar shows $p_{\rm dust}$ in percentage (0--10\%). Only pixels with
    FIR total intensity detected above $3\sigma_I$ ($\sigma_I = 0.04$\,Jy\,pixel$^{-1}$) are shown, which clips the noise-dominated low-intensity areas. In both panels, white contours indicate the SFR surface-density levels of 0.10, 0.20, 0.30, 0.40, 0.50, delineating the most
    actively star-forming regions of the galaxy, tracing the optical spiral arms and several bright star-forming complexes.}
    \label{fig:dop}
\end{figure*}

Figure~\ref{fig:dop} (left) shows the synchrotron polarization fraction of NGC\,6946 at $\lambda = 6$\,cm. High polarization fractions are found in between the spiral arms
reaching values of 30--50\%. These are the regions which host the most ordered magnetic fields in the galaxy, forming the so-called ``magnetic arms'', where the large-scale
dynamo can act more efficiently in the relative absence of strong turbulence \citep{Beck+1996}. The high $p_{\rm syn}$ values in these regions confirm that the ordered
magnetic field is coherent on scales larger than the synthesized beam and is only weakly affected by Faraday depolarization. This morphology is consistent with the classical
picture of NGC\,6946 as a galaxy with strongly contrasted arm and interarm magnetic structures \citep{Beck+1996, Beck2007}.

A pronounced anti-correlation between the synchrotron polarization fraction and the SFR contours is evident in Fig.~\ref{fig:dop} (left). Regions within or near the highest
SFR contour levels exhibit very low $p_{\rm syn}$, typically below 10\%. This anti-correlation can be due to both low polarized emission $PI$ and intense synchrotron
radiation in star-forming regions (see Fig.~\ref{fig:ngc6946_maps}). In these regions, strong depolarization can be caused by the effects of star-formation-driven
turbulence. The injection of kinetic energy by supernovae shocks and stellar winds increases the turbulence (and strength of the turbulent magnetic field component), as well as
the thermal electron density, resulting in both beam and internal Faraday depolarization. These mechanisms reduce the observed $p_{\rm syn}$ in star-forming regions.
Similar effects have been described in detail in previous studies of the magnetic field structure of NGC\,6946 \citep{Beck2007, Tabatabaei2013}.

Figure~\ref{fig:dop} (right) shows the resulting map of $p_{\rm dust}$ at $154\,\mu\mathrm{m}$ in NGC\,6946. As for the radial profiles, only pixels with total intensity
detected above $3\sigma_I$ ($\sigma_I = 0.04$\,Jy\,pixel$^{-1}$) are shown, since at lower intensities the ratio $PI/I$ is dominated by noise. The dust polarization
fraction is typically 1--5\% (median $\sim$2\%), with localized peaks reaching $\sim$9\%, well below the empirical maximum polarization fractions of $\sim$20\%
observed in the diffuse interstellar medium by \citet{Planck2015}. The localized peaks indicate regions with a comparatively ordered magnetic field component in the
plane of the sky.

The map reveals that regions enclosed by the highest SFR contours exhibit the lowest polarization fractions, often approaching $p_{\rm dust} \approx 0$. In contrast,
enhanced $p_{\rm dust}$ is found outside the most active star-forming regions. At the angular scales probed here, the polarization fraction primarily reflects the
relative level of magnetic field order versus disorder (e.g., tangled or turbulent components) along the line of sight and within the beam. The increase of $p_{\rm dust}$
outside the main star-forming complexes therefore suggests a higher degree of ordered magnetic fields in these regions, rather than variations in grain alignment
efficiency \citep{Lopez2024}.

A direct comparison between the polarization position angles derived from FIR and radio data provides a more robust test of the magnetic field structure across ISM phases. In
particular, regions with high $p_{\rm dust}$ (e.g., $\gtrsim 5\%$) are expected to trace well-ordered magnetic fields, and their polarization angles can be directly compared with
those derived from synchrotron emission to assess the coherence of the magnetic field between the cold and diffuse components of the ISM.

The spatial variations of $p_{\rm dust}$ across the galaxy thus trace changes in the relative level of ordered-to-random magnetic field components. The discussion here focuses
on the location of low and high $p_{\rm dust}$ regions within the galaxy, rather than on a direct comparison between dust and synchrotron polarization fractions. Because
the intrinsic polarization mechanisms differ, and because dust polarization fractions can reach empirical upper limits of $\sim$20\% in the diffuse ISM
\citep{Planck2015}, while synchrotron polarization fractions are governed by different physical constraints, the absolute polarization levels of dust and synchrotron
emission are not directly comparable. Grain alignment efficiency and magnetic field ordering are both suppressed in high-SFR environments but persist in the more quiescent
interarm regions. This behaviour is analogous to the arm--interarm contrast seen in radio synchrotron polarization, although the absolute values and spatial patterns
differ due to the distinct emission mechanisms.

\begin{figure}
    \centering
    \includegraphics[width=0.45\textwidth]{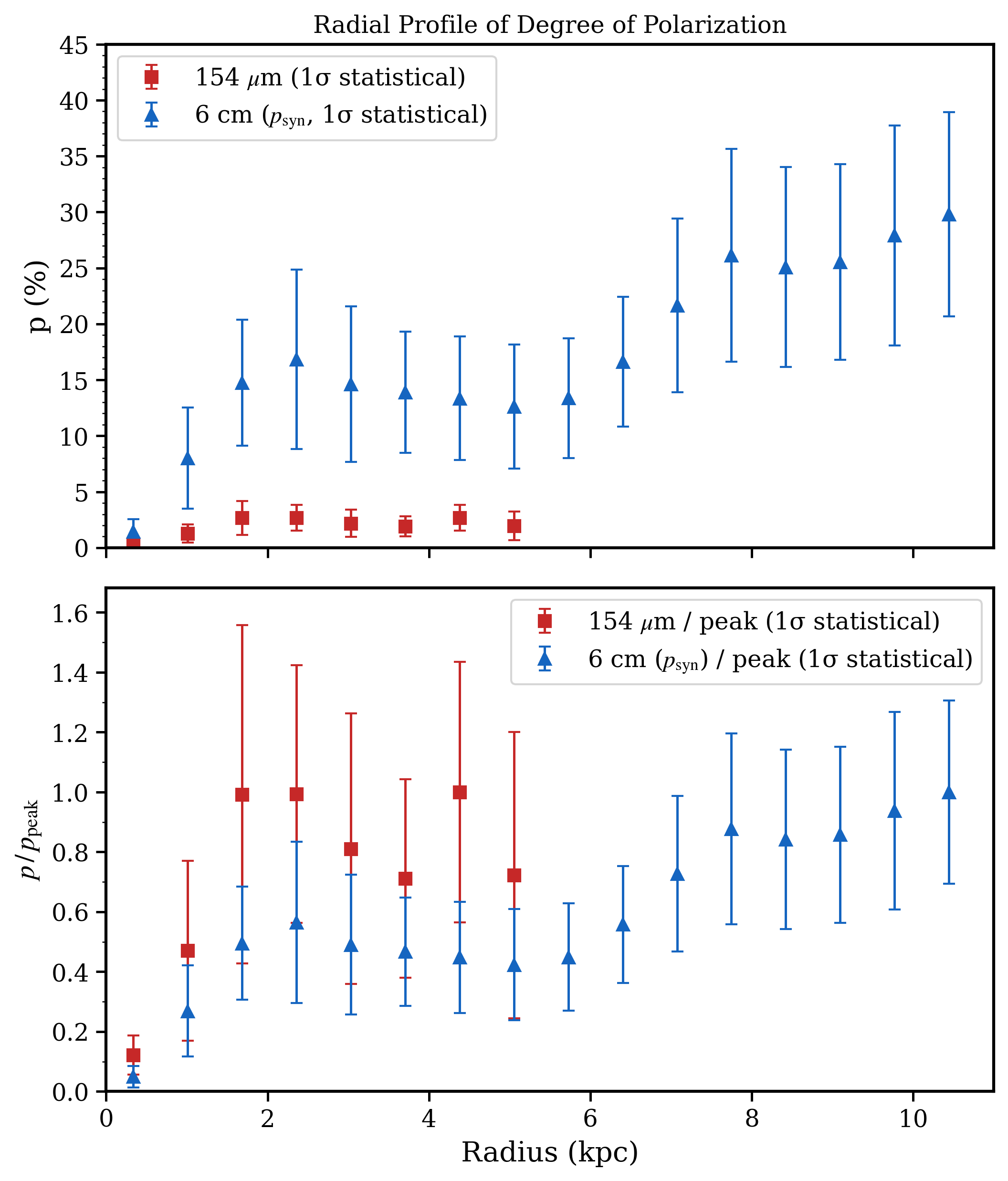}
    \caption{Top: Radial profiles of the fraction of synchrotron polarization $p_{\rm syn}$ (blue triangles) and fraction of dust polarization $p_{\rm dust}$ (red
squares) in NGC\,6946. Error bars indicate the statistical uncertainty (standard deviation of $p$ within each radial bin); systematic uncertainties are not included.
Only pixels with total intensity above $3\sigma_I$ are included, which restricts the dust profile to $R \lesssim 5.4$\,kpc --- smaller than the SOFIA/HAWC+ coverage shown
in Fig.~\ref{fig:ngc6946_maps} and than the radio coverage. Bottom: as top, but with each tracer normalized to its own peak value. The y-axis range in each panel
has been set to include all data points with their full error bars.}
    \label{fig:dop_radial_profile}
\end{figure}

Figure~\ref{fig:dop_radial_profile} presents the radial profiles of the fraction of synchrotron and dust polarization. Radial
profiles were computed in concentric deprojected annuli of width equal to the common angular resolution of the maps, $18''$ (corresponding to a physical width of $\sim 0.67$\,kpc at
the adopted distance of 7.72\,Mpc), by averaging the per-pixel polarization fractions within each annulus (excluding blank pixels). To ensure reliable polarization
fractions, only pixels with total intensity detected above the same significance thresholds adopted in Sect.~\ref{sec:radial_structure} are included: $3\sigma_I$ for the
dust data ($\sigma_I = 0.04$\,Jy\,pixel$^{-1}$; Table~\ref{tab:NGC~6946_data}), restricting the $p_{\rm dust}$ profile to $R \lesssim 5.4$\,kpc (smaller than both the
SOFIA/HAWC+ coverage shown in Fig.~\ref{fig:ngc6946_maps} and the radio extent of $\sim 10$\,kpc), and $8\sigma_I$ for the radio total intensity (Sect.~\ref{sec:radial_structure}).
For the synchrotron polarization fraction, pixels are additionally required to have polarized intensity above $3\sigma_{PI}$ (Sect.~\ref{sec:radial_structure}) and
nonthermal (synchrotron) intensity $I_{\rm syn}$ above $3\sigma_{\rm syn}$ ($\sigma_{\rm syn} = 35\,\mu$Jy\,beam$^{-1}$; Appendix~\ref{TNT}), to avoid noise-dominated
$p_{\rm syn}$ estimates where the synchrotron intensity approaches the detection limit.
The error bars in Fig.~\ref{fig:dop_radial_profile} represent purely statistical uncertainties (the standard deviation of $p$ values within each annulus); systematic uncertainties such as calibration errors or residuals from the thermal/nonthermal separation are not included.

The synchrotron polarization fraction, which primarily reflects the coherence of the large-scale magnetic field ($B_{\rm ord}$), remains relatively high across most of the
disk: $p_{\rm syn}$ is strongly suppressed to a few per cent in the central kiloparsec, holds at $\sim 13$--$17\%$ across the mid disk ($R \approx 2$--$6$\,kpc), and rises to
$\sim 25$--$30\%$ beyond $R \approx 7$\,kpc. The suppression in the inner, actively star-forming disk is indicative of
enhanced depolarization due to a stronger turbulent magnetic field component ($B_{\rm ran}$) and intrinsic Faraday depolarization in star-forming regions
\citep{Beck2015}, while the outer increase reflects the dominance of the ordered interarm field where turbulent driving by star formation is weak.

The dust polarization fraction $p_{\rm dust}$ is systematically lower across all radii, spanning $\sim 0.3\%$ to $\sim 3\%$: it is lowest in the innermost bin
($\sim 0.3\%$) and remains approximately constant at $\sim 2$--$3\%$ between $R \approx 1.5$ and $5$\,kpc, with no systematic radial trend. Because synchrotron and dust emission have
different intrinsic maximum polarization fractions, the absolute values of $p_{\rm syn}$ and $p_{\rm dust}$ should not be compared directly. Instead, the more meaningful
comparison is how each tracer reflects the relative contributions of the ordered and random magnetic field components in the different ISM phases. In this context, the low
values of $p_{\rm dust}$ suggest that, within the dusty ISM, the random-field contribution is comparatively stronger and/or grain alignment is less efficient,
particularly in dense molecular environments \citep{SALSAIV}. Overall, the synchrotron and dust polarization fractions differ both in absolute level and in radial
behaviour: $p_{\rm syn}$ remains systematically higher, tracing a coherent, large-scale ordered field in the diffuse ionized ISM, while
$p_{\rm dust}$ is systematically lower and, although radially flat within the uncertainties, shows strong local (pixel-to-pixel) variations in the map
(Fig.~\ref{fig:dop}), reflecting a more turbulent and less ordered magnetic field in the cold, dense ISM phase (quantified by the pitch-angle dispersion
$\sigma_\psi$ in Sect.~\ref{general}). To compare the radial \emph{shape} of the two profiles despite their $\sim$10x difference in absolute level, we also
normalize each tracer to its own peak value (Fig.~\ref{fig:dop_radial_profile}, bottom panel); the resulting trends are similar to those seen in the top panel, but the
radial variation of $p_{\rm dust}$ is rendered more clearly visible once the $\sim$10x offset in absolute level between the two tracers is removed.

\section{Discussion}
\label{sec:discussion}

{As tracers of the magnetic field, the distributions of the FIR and synchrotron polarization presented earlier are compared more quantitatively in different galactic regions
in this section. We investigate any correlations with the SFR and gas surface densities and discuss the origin of the FIR polarization taking into account relevant physical
parameters. In addition, we compare our results in NGC\,6946 with similar studies in M~51.}

\begin{figure}
    \centering
    \includegraphics[width=0.45\columnwidth]{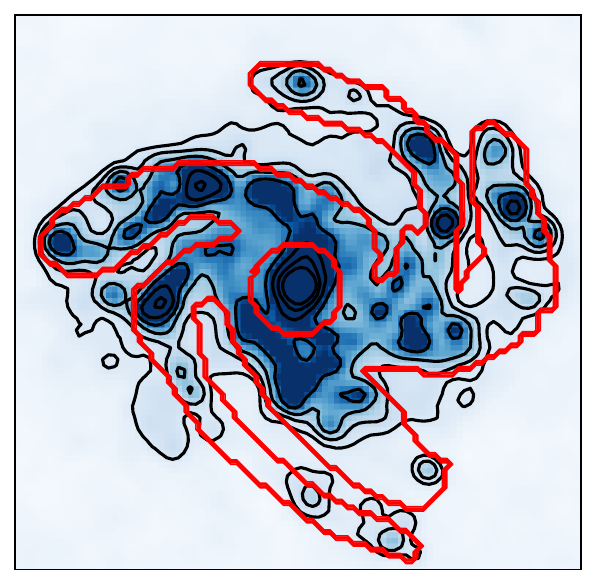}
    \vspace{3mm} 
    \includegraphics[width=0.45\columnwidth]{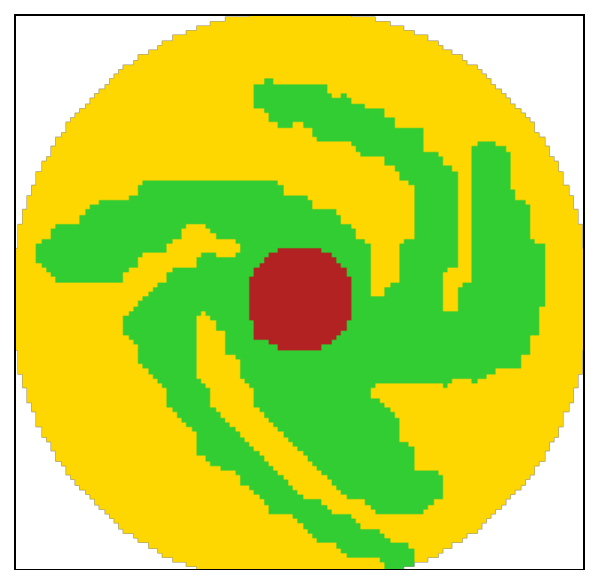}
    \caption{The mask used to distinguish the central region(\textit{red}), arm (\textit{green}), and interarm (\textit{yellow}) regions of NGC\,6946. The left panel shows
    the mask contours overlaid on the PACS 70\,$\mu$m map, while the right panel displays the mask itself.}
    \label{fig:regionalmask}
\end{figure}

{To study the properties in different regions of NGC\,6946, 
we use a mask, originally introduced by \cite{Bigiel+2020}, which is based on the PACS 70\,$\mu$m map and distinguishes spiral arms, inter-arms, and
central region
(inner 48\as or 1.4\,kpc region). To account for the larger field of view provided by SOFIA, we modified this mask by expanding its outer edge to cover a larger area in the
inter-arm regions  (Fig.~\ref{fig:regionalmask}). However,
as we are restricted to 
the sensitivity of the FIR maps (Sect.~\ref{general}),
 the inter-arm region are left with only few data points, preventing statistically reliable analysis. Therefore, we focus on the spiral arms and the central region of
 NGC\,6946.}

\subsection{Comparison of Dust and Synchrotron Polarization}

\begin{figure}[htbp]
\centering
\includegraphics[width=0.45\textwidth]{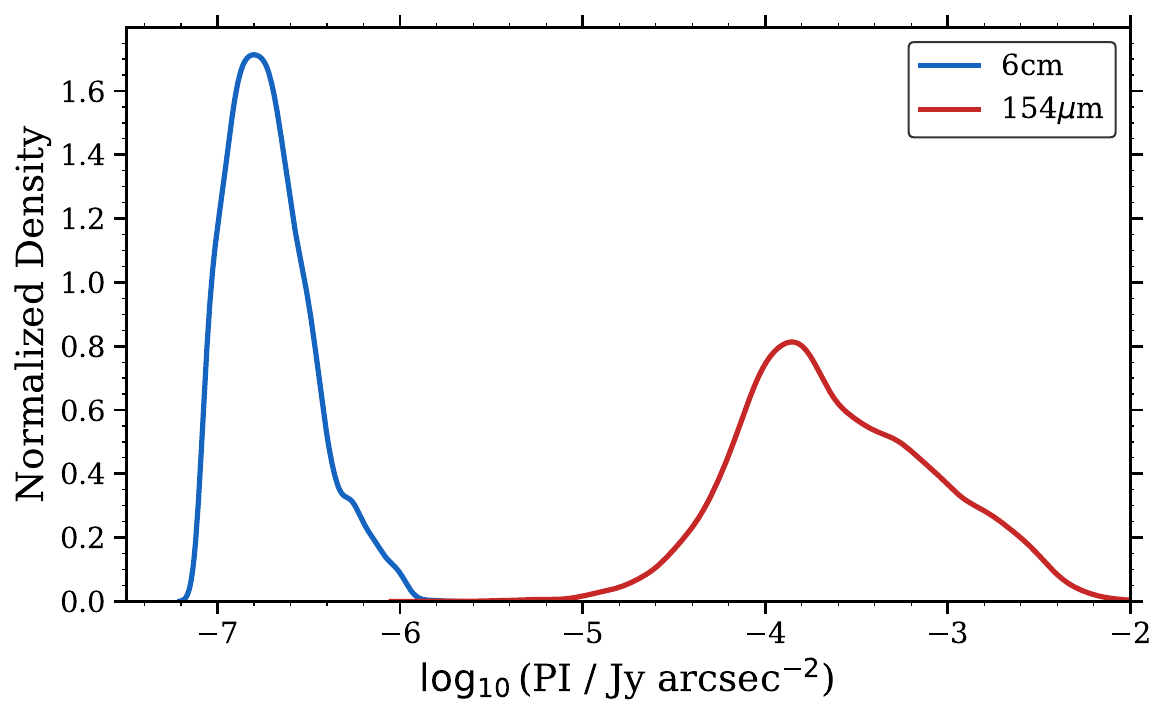}
\caption{Kernel density estimate (KDE) of polarized intensity (PI) distributions at 6\,cm (blue) and 154\,$\mu$m (red) in NGC\,6946. The x-axis shows
$\log_{10}(\mathrm{PI})$ in $\mathrm{Jy\ arcsec^{-2}}$, highlighting the
narrower radio distribution compared to the
broader FIR one.}
\label{fig:kde_pi}
\end{figure}

The kernel density estimate (KDE) plot in Figure~\ref{fig:kde_pi} illustrates the distributions of polarized intensity (PI) values at 6\,cm (blue curve) and 154\,$\mu$m
(red curve) across NGC\,6946, revealing distinct differences in their typical intensities and spreads. The radio PI at 6\,cm peaks at approximately
$\log_{10}(\mathrm{PI}) \approx -6.5$ (corresponding to $\sim 6 \times 10^{-7}\ \mathrm{Jy\ arcsec^{-2}}$), with a narrow distribution extending primarily from $-7$ to $-5$.
The FIR PI at 154\,$\mu$m peaks at $\log_{10}(\mathrm{PI}) \approx -4.2$ (corresponding to $\sim 6 \times 10^{-5}\ \mathrm{Jy\ arcsec^{-2}}$), with a broader distribution
from $-5.5$ to $-3$.

\begin{figure*}[htb!]
    \centering
    \includegraphics[width=0.48\textwidth]{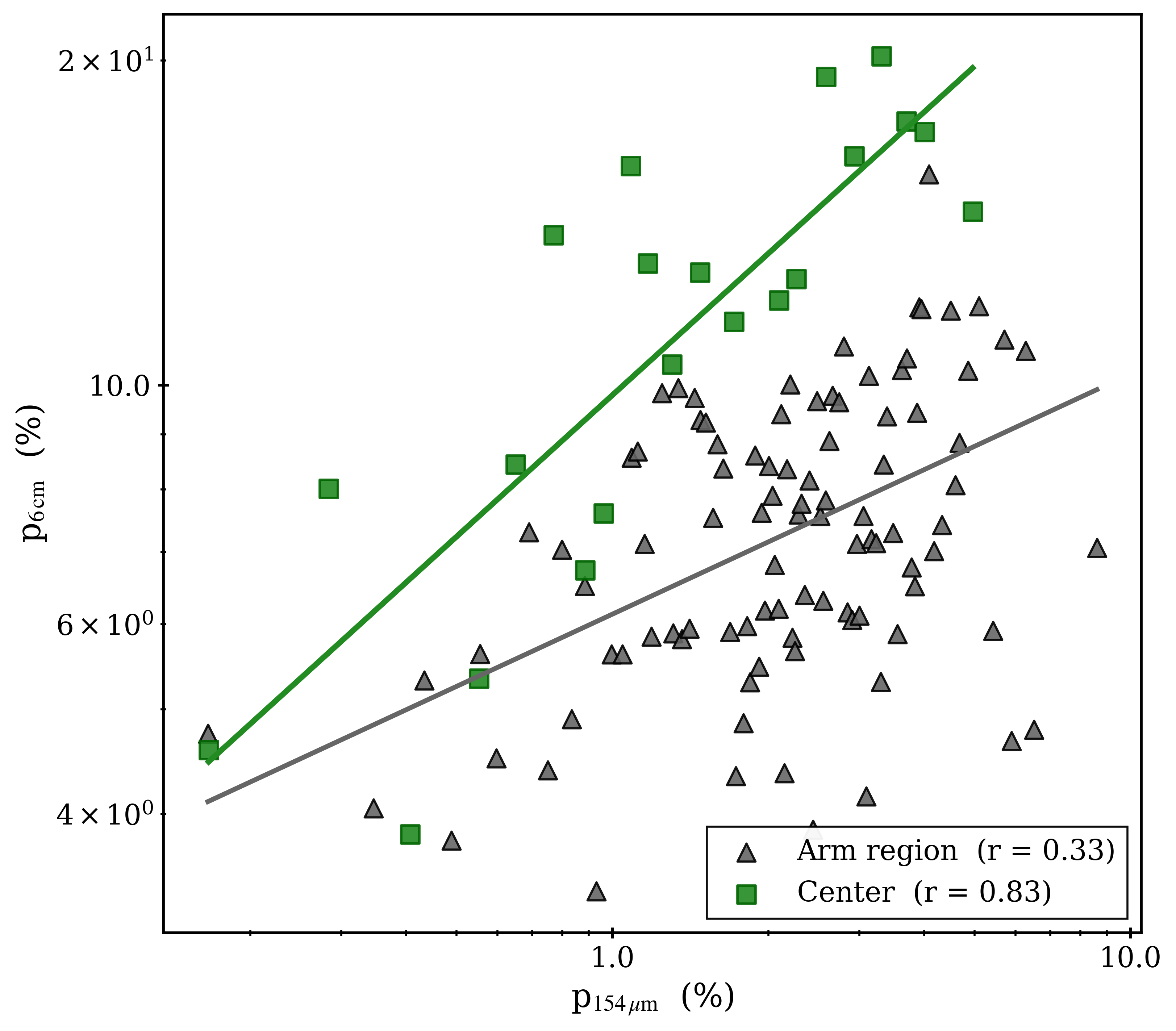}
    \hfill
    \includegraphics[width=0.48\textwidth]{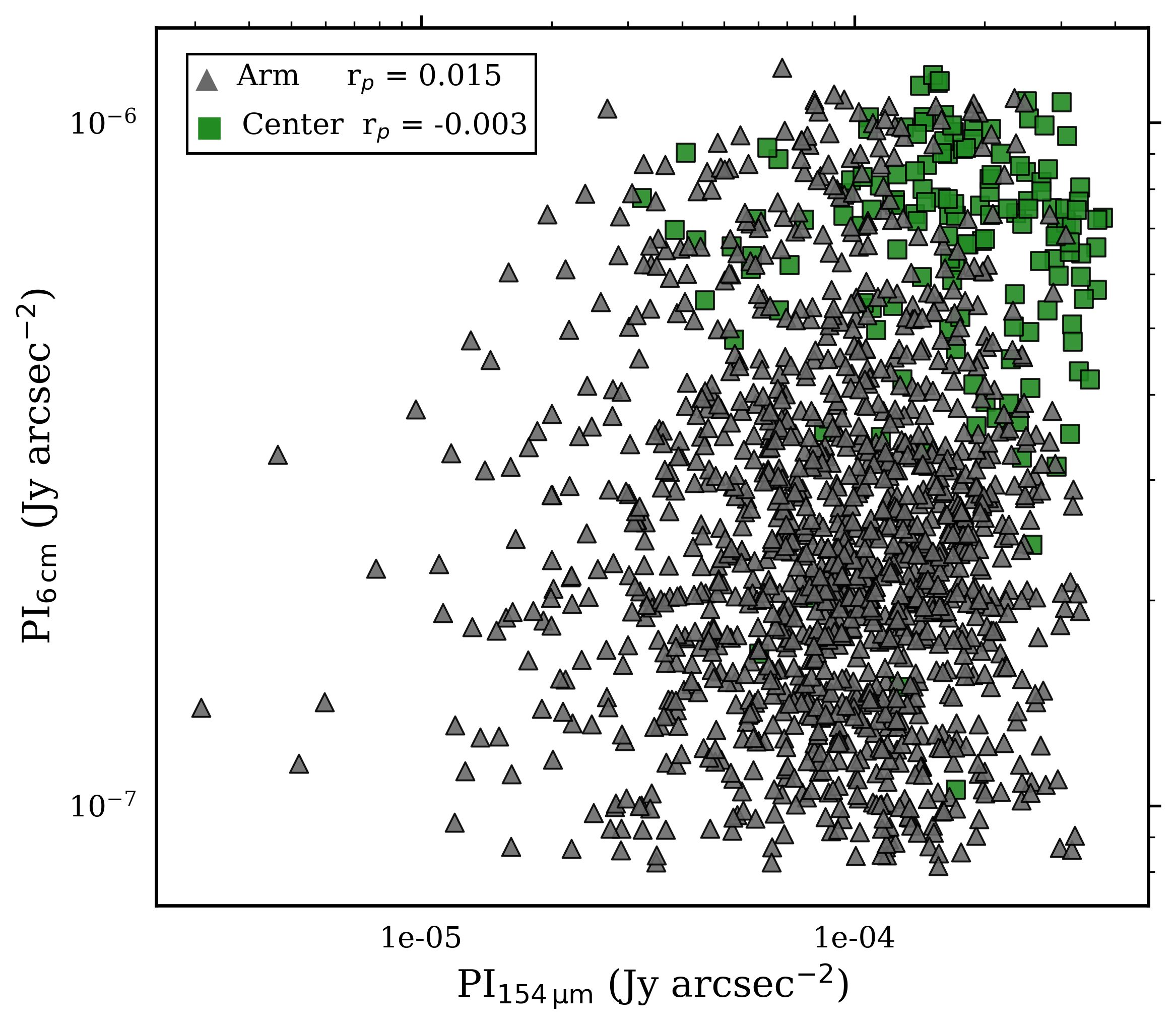}
    \includegraphics[width=0.48\textwidth]{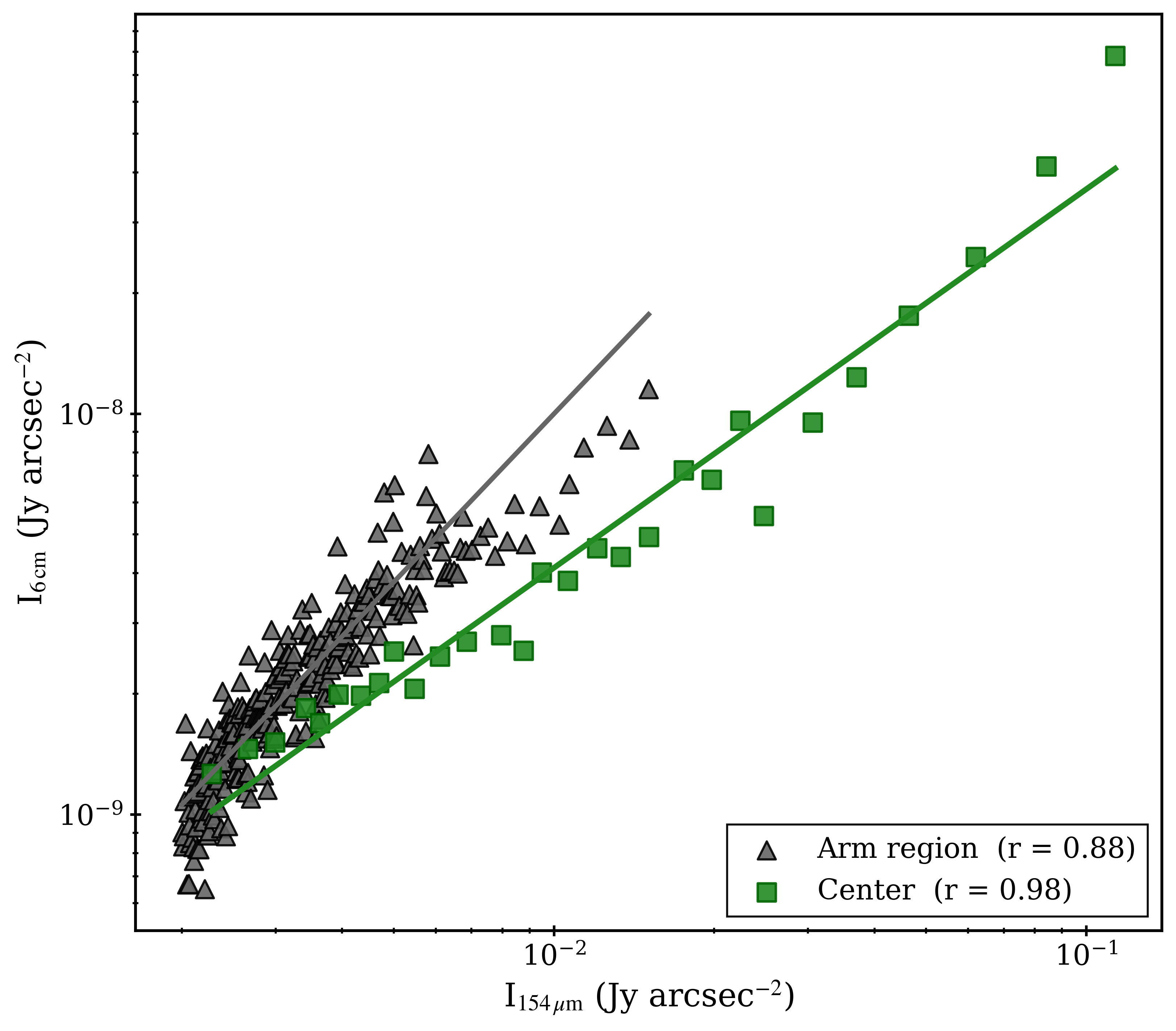}
    \caption{Pixel-by-pixel comparison of the polarization and intensity properties at 154\,$\mu$m and 6\,cm in NGC\,6946. The left panel shows the relation between the
    polarization fractions at 154\,$\mu$m and 6\,cm, the right panel presents the comparison between the polarized intensities at 154\,$\mu$m (dust emission) and 6\,cm
    (synchrotron emission), and the bottom panel shows the correlation between the total 6\,cm radio continuum surface brightness, $I_{6\,\mathrm{cm}}$, and the 154\,$\mu$m
    far-infrared surface brightness, $I_{154\,\mu\mathrm{m}}$. Gray triangles correspond to spiral-arm regions, while green squares denote the central region. Solid lines
    indicate the best-fit relations where applicable.
}
    \label{fig:dop_pi}
\end{figure*}

Studying the correlation between the dust and synchrotron tracers is a first step toward investigating any connection between the magnetic field in the neutral and ionized gas components. As shown in Fig.~\ref{fig:dop_pi}, no correlation is found between
the polarized intensities (Pearson correlation coefficients are $r \approx 0$ in both the spiral arms and the central region). On the other hand, the fractional polarizations
are correlated: the spiral-arm regions (gray triangles) show a weak-to-moderate correlation ($r \approx 0.33$), while the central-region points (green squares) exhibit a relatively
tight relation ($r \approx 0.83$).

Figure~\ref{fig:dop_pi} presents the pixel-by-pixel comparison between the polarized intensities and fractional polarizations at 154$\mu$m (dust emission) and 6 cm (synchrotron emission).

The correlation in fractional polarization is hence because of the correlation of the total intensities. The lack of correlation in PI can refer to different (1) distributions of relativistic particles and dust grains, or (2) intervening physical processes. The first possibility is not confirmed by the observed tight FIR--synchrotron correlation (Fig.~\ref{fig:dop_pi}, panel c), which even persists in quiescent parts of the ISM in this galaxy \citep{Tabatabaei+2013}, reflecting a coupling between the magnetic field, cosmic rays, and the dusty ISM. The second possibility is more relevant, as explained in the following.

For instance, the synchrotron PI is affected by turbulence due to e.g., Faraday dispersion, while the dust PI depends more critically on other physical parameters such as radiation field and gas density (see Sects.~\ref{sec:gas_phases} and \ref{sec:fir_origin}).
This introduces considerable scatter in the PI comparison, particularly because the total radio and FIR emissions correlate tightly due to their shared
dependence on star formation rate and gas density \citep{Tabatabaei2013}, but the polarized subsets are more sensitive to field ordering and depolarization effects.
Normalizing by total intensity to obtain p isolates the efficiency of magnetic field ordering and dust grain alignment, revealing a more consistent relationship between the
two wavelengths as they reflect coupled physical processes across ISM phases. For instance, the scattered PI points in the arms arise from positional mismatches between dust
lanes (high FIR PI but variable total intensity) and synchrotron-emitting regions (enhanced by cosmic rays in star-forming areas), but their p values align better due to
shared underlying field topology. In the center, the near-zero PI correlation reflects amplified unpolarized components from dense gas and stronger depolarization, yet the
high p correlation indicates coherent ordering persisting across phases.

Physically, the observed p correlation provides evidence for coupling between the magnetic fields in the ionized ISM (traced by synchrotron) and in the
neutral/molecular ISM (traced by dust), where an increase in the alignment/ordering of the magnetic field in the neutral/molecular phase is generally accompanied by increased
ordering in the ionized phase. However, the radio polarization appears highly depolarized overall, contributing to the cloud-like scatter in both p and PI plots, though less
so in p after normalization. The polarization fraction at 154\,$\mu$m reaches only a few per cent in the core region (median $\sim$1\%, up to $\sim$6\%), lower than the high
median p values (e.g., $8.6\% \pm 4.8\%$) reported for similar galaxies \citep{SALSAIV}, consistent with reduced grain alignment efficiency and stronger field tangling in
this dense, actively star-forming environment. The stronger p
correlation in the central region suggests that the magnetic field is more coherent across phases there, likely due to strong shear and compression that organize the field on
scales larger than line-of-sight variations, rather than a mean-field dynamo. In contrast, the weaker correlation in the arms indicates that
additional factors reduce the coupling without implying full decoupling: local star-formation feedback, small-scale turbulence, and complex mixtures of optical depths and
radiation anisotropies act to partially decorrelate the tracers, for example by tangling fields on sub-beam scales or reducing the effective polarization through vector
cancellations.

Several important caveats and physical mechanisms must be considered when interpreting the scatter and slope of the p relation, as these also explain the even weaker PI
trends. Firstly, although the observations are matched to the same resolution, the two tracers sample different volumes along the line of sight: FIR polarization
preferentially reflects conditions in the neutral/molecular gas, whereas synchrotron polarization averages over a larger path length that includes ionized gas from the disk
and diffuse halo; this geometric difference contributes to scatter, as supported by magnetohydrodynamic simulations showing tomographic variations in multi-wavelength
polarimetry \citep{martin2023tomographic}. Secondly, non-thermal radio polarization is affected by Faraday rotation and depolarization (internal and external), which can
reduce the measured radio p independent of the underlying field ordering; these effects are wavelength-dependent and may vary strongly between the arms and center, with
synchrotron p increasing with gas density in the arms (indicating amplification in denser ionized regions) while dust p shows no such correlation \citep{SALSAIV}. Thirdly,
at the scales of these observations (typically $\gtrsim 100$\,pc), radiative alignment torque (RAT) effects on grain alignment are negligible, with the morphology of the
magnetic field (pure geometrical effects, such as tangling and projection) dominating the polarization variations; RAT mechanisms become important only at sub-pc scales in
specific molecular clouds and star-forming regions, as demonstrated by Planck observations of Galactic fields and detailed studies of spiral galaxies \citep{Planck2015, Lopez2024}. Finally, observational effects like beam dilution, differing signal-to-noise masks, and residual calibration errors or bias in debiased
polarization estimates will broaden both relations.

Despite these caveats, the joint FIR--radio p correlation, contrasted with the weak PI scatter, is a powerful diagnostic of multi-phase magnetism. The strong central
coupling (high $r$) argues for a coherent magnetic topology that links the ionized and neutral/molecular ISM in the central kpc, whereas the moderate arm coupling points to a
regime where small-scale turbulence and localized physics (star formation, shocks) increasingly influence the observed polarization fractions.

Quantifying the above statements in detail would require three additional analyses that are beyond the scope of the present study. First, separating
turbulence-driven depolarization from purely geometrical (tangling and projection) effects \citep{SALSAIV} would need independent constraints on the three-dimensional
field structure, such as a resolved rotation-measure map, beyond the two-dimensional polarization maps used here. Second, identifying the specific beam- and
line-of-sight-depolarization contributions \citep{SALSAI} would require higher-resolution or multi-frequency observations than are currently available for this galaxy.
Third, dedicated radiative-transfer modeling of the grain-alignment physics would require additional multi-wavelength radiative-transfer calculations beyond the
empirical comparison presented here. We leave these three analyses for a dedicated follow-up study.

A related question is whether the amplitude of the turbulent magnetic field, $B_{\rm turb}$, could be estimated on the scales probed here, given that the
synchrotron polarization fraction is more sensitive to the ordered field while the FIR polarization fraction (and its angular dispersion, Sect.~\ref{general}) is more
sensitive to the turbulent component. Because radio and FIR trace physically distinct ISM phases along the line of sight (diffuse ionized gas versus dense
neutral/molecular gas), observed here at the same angular resolution, we would in fact expect the two tracers to yield different $B_{\rm turb}$ estimates; measuring
this difference, rather than being an obstacle, is precisely what would make such a joint analysis worthwhile. In principle, this is possible: the classical
Chandrasekhar--Fermi method \citep{ChandrasekharFermi1953} relates $B_{\rm turb}$ to the polarization angle dispersion $\sigma_\psi$, the line-of-sight gas density
$\rho$, and the gas velocity dispersion $\sigma_v$ along the same line of sight ($B_{\rm turb} \propto \sqrt{4\pi\rho}\,\sigma_v/\sigma_\psi$), and has been applied to
both synchrotron and dust polarization angle maps individually in the Galactic ISM \citep{Hildebrand2009, Houde2009}. Applying it to our own $\sigma_\psi$ measurements
(Sect.~\ref{general}) together with the CO and H\,\textsc{i} line widths available for NGC\,6946 (Sect.~\ref{sec:data}) could yield $B_{\rm turb}$ estimates for each
tracer separately. We do not attempt this here, for two reasons. First, the Chandrasekhar--Fermi method assumes isotropic turbulence in approximate energy
equipartition with the magnetic field, an assumption that is harder to justify in the multi-phase, star-formation-driven ISM of a whole galaxy disk than in the resolved
individual clouds it was originally developed for. Second, Faraday rotation inflates the observed radio angular dispersion relative to the true one (Sect.~\ref{dp}),
which would bias any Chandrasekhar--Fermi estimate from the synchrotron data high unless explicitly corrected for. A careful joint Chandrasekhar--Fermi analysis of both
tracers, accounting for these effects, is a promising direction for future work but is beyond the scope of the present paper. We note that $B_{\rm turb}$ can also be
estimated independently, and with higher accuracy than via the Chandrasekhar--Fermi method, from the unpolarized synchrotron intensity, which isolates the isotropic
turbulent field component directly \citep{Beck2025}.

\subsection{Correlations with star formation rate}

Theoretically, star formation activity can amplify magnetic fields by injecting turbulence into the ISM, for example through small-scale dynamo action \citep{Gressel2008}.
This is supported by the tight correlation between the equipartition magnetic field strength traced by synchrotron emission and the SFR, observed both locally
\citep{Tabatabaei2013, Hassani2022, Nasirzadeh2024} and globally in galaxies \citep{Chyzy2008, Heesen2014, Tabatabaei2017}. However, any correlation between SFR and polarized
emission can be complicated by the turbulent nature of star formation activity, as indicated by synchrotron polarization studies \citep{Tabatabaei2013, Nasirzadeh2024}. In
the following, we compare the impact of the SFR on polarization signals observed in the FIR and radio regimes in NGC\,6946.

\begin{figure*}
\centering
\includegraphics[width=0.48\linewidth, height=0.6\textheight]{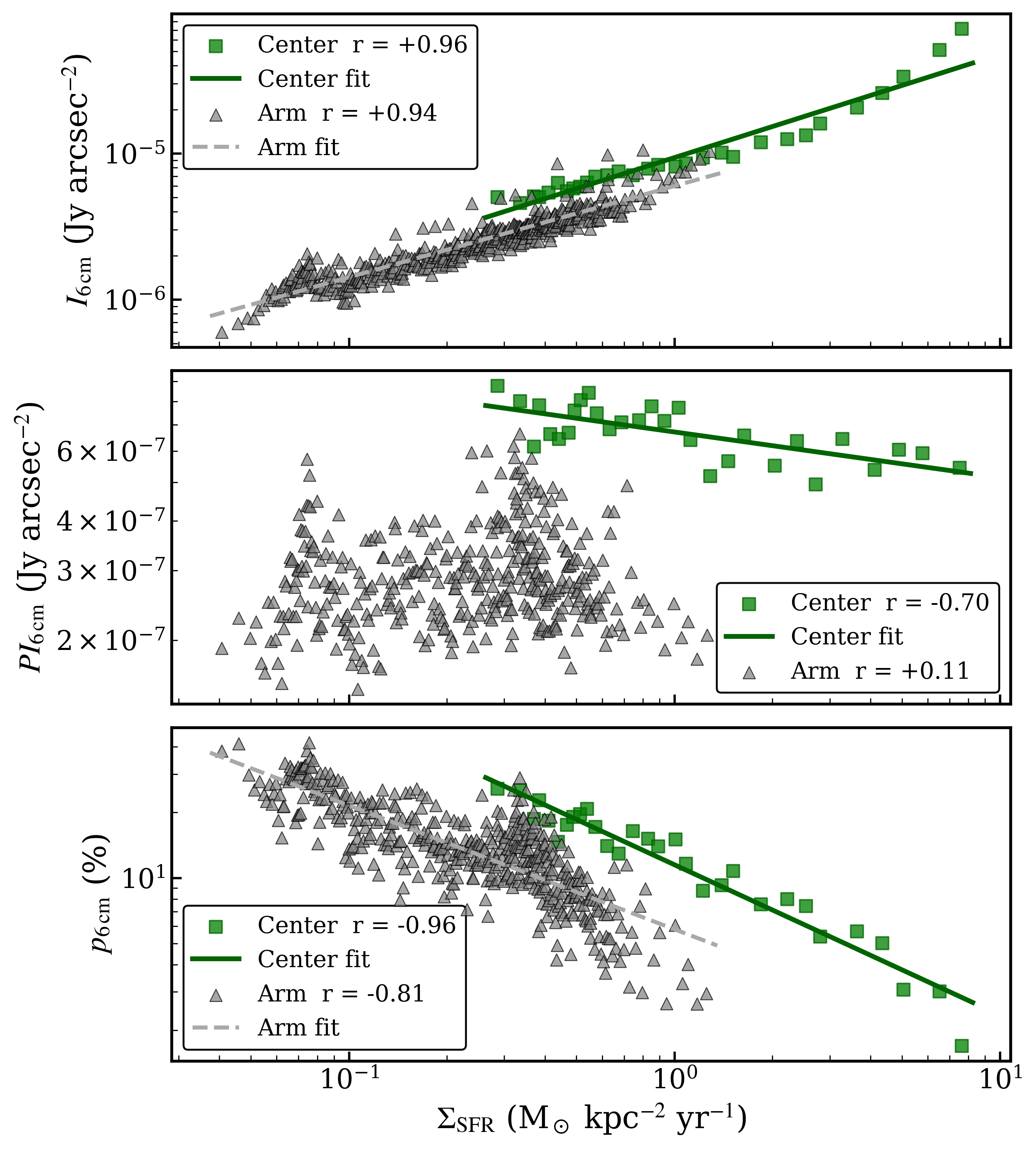}
\hfill
\includegraphics[width=0.48\linewidth, height=0.6\textheight]{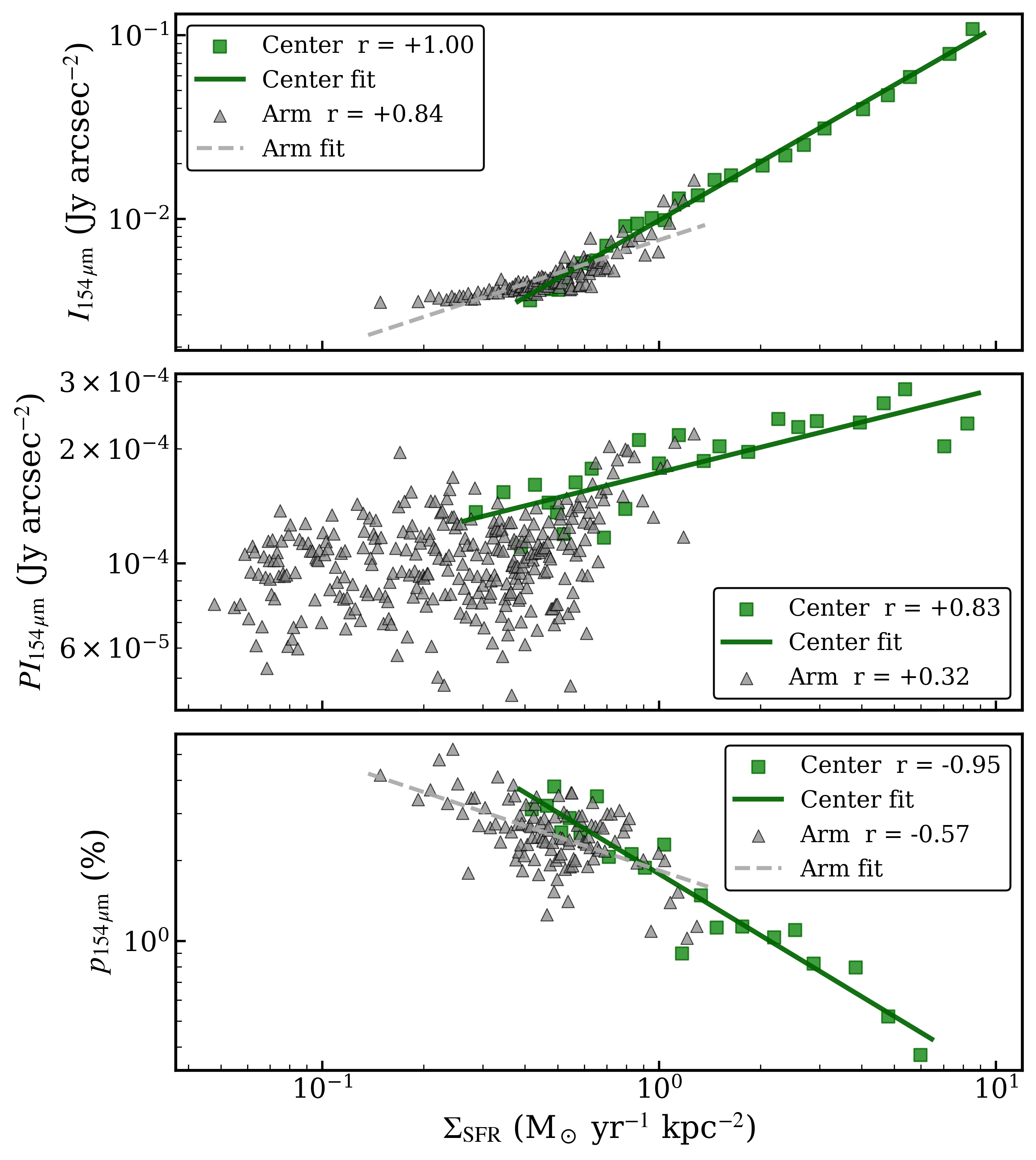}
\caption{{The radio 6\,cm (left) and FIR 154\,$\mu$m (right) emission vs the star formation rate (SFR) in NGC\,6946,
for the total synchrotron intensity (top), polarized intensity (middle), and polarization fraction (bottom) in the spiral arms (triangles) and central region (squares).}}
\label{fig:FIR+radio-SFR}
\end{figure*}

Figure~\ref{fig:FIR+radio-SFR} compares the 6\,cm (radio) and 154\,\micron\ (FIR) polarized emission and polarization fraction as functions of the star formation rate surface
density, $\Sigma_\mathrm{SFR}$, for the central region and spiral arms of NGC\,6946.
The left and right columns show measurements at 6\,cm and at 154\,\micron, respectively, with total synchrotron intensity $I$, polarized intensity $PI$, and polarizatio fraction $p$ presented in separate rows.
Data points differentiate the central region (green squares) from the spiral arms (grey triangles), with binned averages and power-law fits (solid lines), along with the
corresponding correlation coefficients $r$.

At 6\,cm, the emission is dominated by synchrotron radiation from cosmic-ray electrons interacting with the magnetic field.
The total synchrotron intensity $I_{6\,\mathrm{cm}}$ increases nearly linearly with $\Sigma_\mathrm{SFR}$, with slopes of $\sim 0.9$ in both the central region and the arms
($r \approx +0.9$).
This behavior is consistent with the established radio--SFR relation \citep[e.g.,][]{Condon1992}, in which higher $\Sigma_\mathrm{SFR}$ enhances cosmic-ray injection and the
total magnetic field strength, producing an approximately linear scaling.

The polarized intensity $PI_{6\,\mathrm{cm}}$ shows a substantially weaker dependence on $\Sigma_\mathrm{SFR}$.
In the central region, $PI_{6\,\mathrm{cm}}$ displays a moderate
negative correlation (slope $\sim -0.6$ -- $-0.7$, $r \sim -0.7$),
while in the arms
no dependence is visible.
This
reflects competing effects: field ordering by large-scale streaming motions and gas compression \citep{Beck2015}, versus strong depolarization caused by turbulence, beam
averaging, and line-of-sight variations.
In regions with elevated $\Sigma_\mathrm{SFR}$, increased turbulence and Faraday effects typically dominate, reducing the observed polarized signal.

The polarization fraction $p_{6\,\mathrm{cm}}$ exhibits a strong anti-correlation with $\Sigma_\mathrm{SFR}$, with slopes of $\sim -0.8$ to $-1.0$ ($r = -0.96$ and $-0.81$ in the
central region and arms, respectively).
This trend is consistent with enhanced depolarization in more actively star-forming regions, where increased turbulence and small-scale magnetic field fluctuations reduce the
ordered-to-total field ratio along the line of sight.
Faraday depolarization is
important at 6\,cm, amplifying this effect in denser environments \citep{Fletcher2011}.

At 154\,\micron, the emission traces thermal radiation from dust grains, with polarization arising from radiative-alignment torques (RATs) acting on non-spherical grains
\citep{Andersson2015}.
The total intensity $I_{154\,\micron}$ scales nearly linearly with $\Sigma_\mathrm{SFR}$ in the central region (slope $\sim 1.1$, $r \approx +1.0$) and more shallowly in the
arms (slope $\sim 0.6$, $r \approx +0.84$), as expected  from FIR–SFR relations where dust heating is tightly linked to recent star formation \citep{Kennicut2012}.

The polarized intensity $P_{154\,\micron}$ shows a positive correlation with $\Sigma_\mathrm{SFR}$, stronger in the central region (slope $\sim 0.2$, $r \approx +0.83$)
and weaker in the arms (slope $\sim 0.1$, $r \approx +0.32$).
Stronger densities of the radiation field
at larger $\Sigma_\mathrm{SFR}$ enhance RAT alignment efficiency, increasing the intrinsic polarized emissivity.
However, line-of-sight field tangling, unresolved cloud structure, and increasing optical depth partially suppress the correlation, particularly in the arms.

The polarization fraction $p_{154\,\micron}$ decreases with increasing $\Sigma_\mathrm{SFR}$, with slopes of $\sim -0.8$ (central region) and $\sim -0.4$ (arms), and
correlation coefficients of $r = -0.95$ and $-0.57$, respectively.
Unlike the radio case, FIR emission is unaffected by Faraday rotation; the decline therefore reflects increased turbulent field disorder, higher column density, and greater
line-of-sight complexity in regions of enhanced star formation \citep{planck2018xi}.

Despite the different emission mechanisms, both radio and FIR data show similar empirical behavior:
\[
p \propto \Sigma_\mathrm{SFR}^{-0.7 \text{ to } -1.0}.
\]
This suggests that the dominant factor reducing polarization in high-$\Sigma_\mathrm{SFR}$ regions is the decrease in large-scale field coherence driven by turbulence, rather
than tracer-specific mechanisms such as Faraday rotation or RAT efficiency.

The polarized intensities show more tracer-dependent differences.
Radio $PI_{6\,\mathrm{cm}}$ is
affected by Faraday depolarization and small-scale field tangling, especially in the arms, while FIR $PI_{154\,\micron}$ is more robust to these effects and correlates more
directly with dust-aligned fields, particularly in the central region.

\subsubsection{Comparison to M51 (Borlaff et al.\ 2021)}

Our results for NGC\,6946 closely parallel those from the SOFIA Legacy Program study of M51 \citep{SALSAI}.
HAWC+ 154\,\micron\ observations of M51 show a decreasing FIR polarization fraction with increasing $\Sigma_\mathrm{SFR}$ in both arms and central region, with slopes broadly
consistent with the $\sim -0.7$ to $-1.0$ range we find in NGC\,6946.
Similarly, radio polarimetry of M51 at 3--6\,cm \citep{Fletcher2011} shows decreasing $p$ toward high-$\Sigma_\mathrm{SFR}$ regions, reflecting the same turbulence-driven
depolarization mechanism.

A key distinction is that M51, as a density-wave grand-design spiral, shows stronger arm–interarm differences: FIR polarization orientations decorrelate from radio
orientations within the arms, likely due to field tangling in compressed molecular gas \citep{SALSAI}.
In contrast, the higher global SFR of NGC\,6946 produces somewhat smoother transitions between arm and central region,
consistent with the shallower arm slopes and weaker differences that we observe.

The similarity
between NGC\,6946 and M51 strengthens the emerging picture from the SALSA Legacy framework \citep{SALSAI}: across spiral galaxies, $\Sigma_\mathrm{SFR}$ is a primary
regulator of observable magnetic field order, with multiwavelength polarimetry providing complementary views of the diffuse and dense ISM phases.

\subsubsection{Offsets between SFR tracers and the 154\,$\mu$m emission}

We note a systematic offset at several locations between the SFR contours and peaks in the 154\,$\mu$m intensity/polarization map. Such displacements
are physically plausible: the 154\,$\mu$m band predominantly traces colder dust heated at the surfaces of molecular clouds, i.e. photon-dominated regions (PDRs;
\citealt{Tielens2005, Hollenbach1999}), which are expected to reside at the interface between molecular gas and the ionized H\,\textsc{ii} regions that dominate H$\alpha$ and
warm dust emission (\citealt{Draine2011}). In the context of spiral arm dynamics, the observed geometry can be interpreted as part of the canonical sequence --- molecular gas
$\rightarrow$ star formation $\rightarrow$ PDRs --- produced by density-wave streaming motions and the finite timescale for dust heating and photo-processing
(\citealt{Egusa2009, Schinnerer2013, Kreckel2018}).

Before attributing the offsets entirely to ISM structure, we verified (i) the astrometric registration of the datasets, (ii) that the effect persists after convolving the
higher-resolution maps to the 18\arcsec\ resolution of the HAWC+ data, and (iii) that only high S/N regions (S/N$_P \!>\!3$) were considered. After these checks the offsets
remain significant at the $\sim$(0.1--0.5)\,kpc level, consistent with expected PDR scales (\citealt{Pabst2022, Rebolledo2021}). We therefore conclude that a substantial
fraction of the 154\,$\mu$m emission arises in PDRs and cloud surfaces adjacent to the young stellar clusters, although contributions from
projection effects and local feedback-driven dust displacement cannot be excluded (\citealt{Lopez2014, Barnes2020}). Future confirmation using high-resolution CO and PAH maps
(e.g., 8\,$\mu$m; \citealt{Sandstrom2010, Chevance2020}) would help to unambiguously identify the PDR locations.


\subsection{Correlations with neutral gas phases}\label{sec:gas_phases}

\begin{figure*}
\centering
\includegraphics[width=0.48\linewidth, height=0.6\textheight]{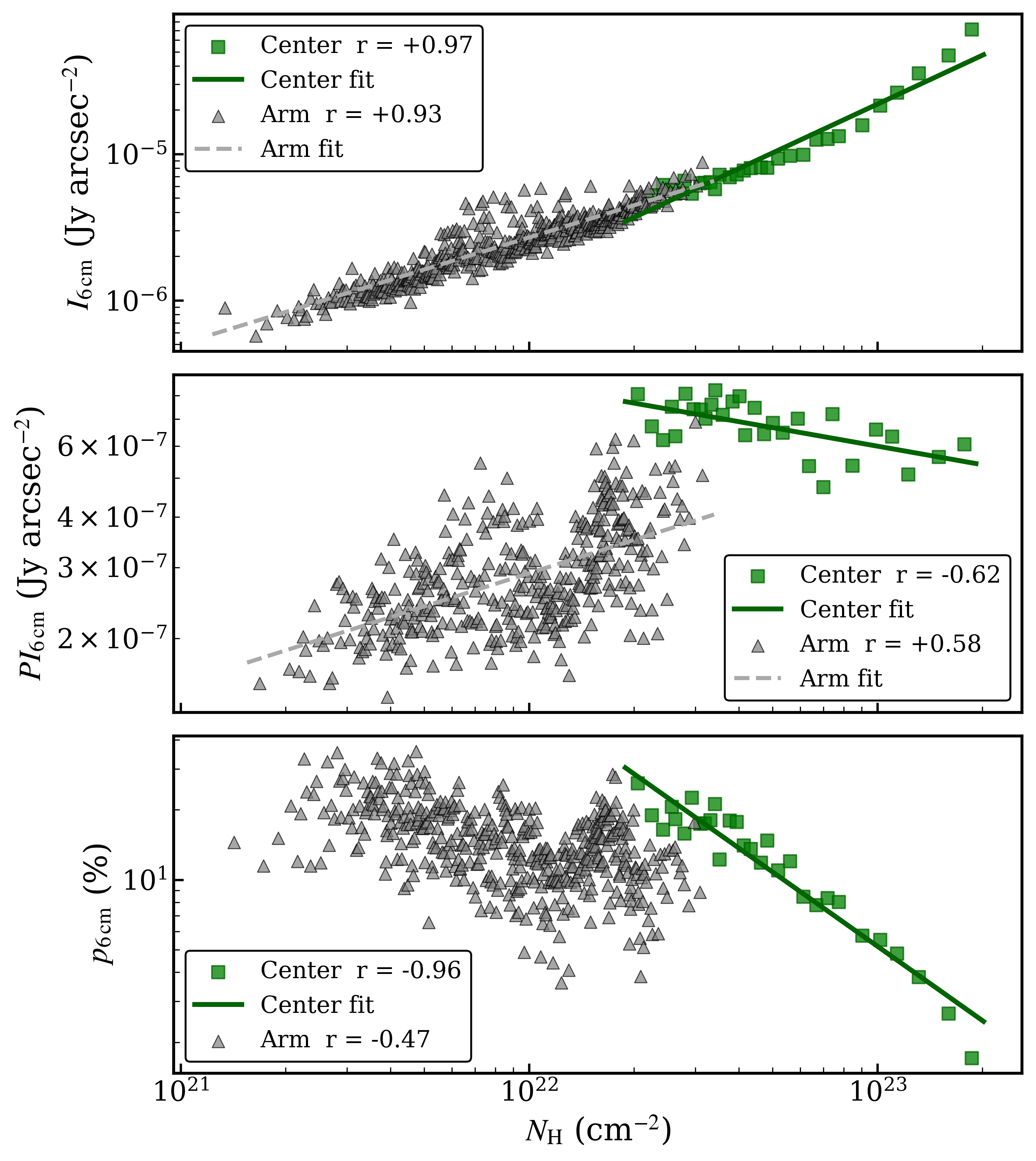}
\hfill
\includegraphics[width=0.48\linewidth, height=0.6\textheight]{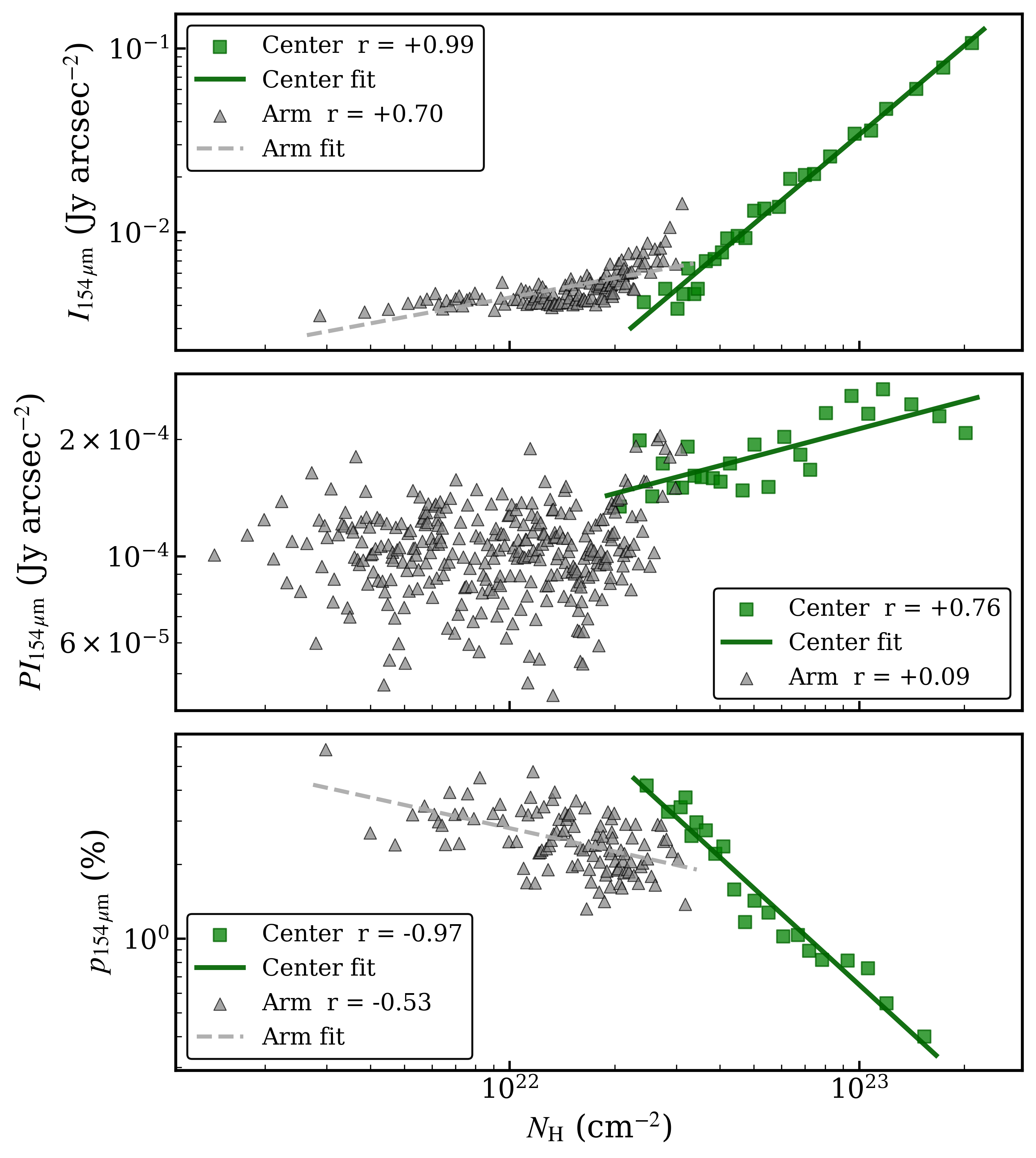}
\caption{Comparison of the 154\,$\mu$m (right) and 6\,cm (left) polarization properties as a function of total gas column density, $N_{\mathrm{H}}$. Each panel shows the
total synchrotron intensity (top), polarized intensity (middle), and polarization fraction (bottom) for the central (green) and arm (grey) regions.
}
\label{fig:FIR+radio-NH}
\end{figure*}

The binned correlations between the far-infrared (154\,\micron) and radio (6\,cm) polarization properties of NGC\,6946 and the total hydrogen column density $N_{\rm H}$
reveal clear and physically meaningful differences between the two tracers.
At 154\,\micron, both the total synchrotron intensity and polarized intensity increase systematically with column density in the central region (Pearson coefficients
$r=+0.99$ and $r=+0.76$), whereas the fractional polarization decreases steeply with $N_{\rm H}$ ($r=-0.97$).
A similar, though weaker, trend is observed in the spiral arms: FIR total intensity rises with $N_{\rm H}$ ($r=+0.70$), but the polarized intensity shows
no correlation ($r=+0.09$), and the fractional polarization declines moderately ($r=-0.53$).
This behavior suggests that in dust-dominated FIR emission, increases in gas column density primarily amplify the dust mass and hence the total and polarized flux while
simultaneously reducing the polarization fraction through depolarization from magnetic field tangling, increased turbulence, or grain-alignment inefficiencies in dense
environments.

The radio 6\,cm emission shows a related but notably distinct pattern.
As expected for synchrotron-dominated radiation, the total synchrotron intensity correlates strongly with $N_{\rm H}$ both in the central region ($r=+0.97$) and in the
arms ($r=+0.93$). The polarized intensity rises with $N_{\rm H}$ in the arms ($r=+0.58$), reflecting the presence of stronger total magnetic fields and larger cosmic-ray
electron densities in gas-rich regions, whereas in the central region it declines with $N_{\rm H}$ ($r=-0.62$).
However, the fractional polarization at 6\,cm decreases sharply in the central region ($r=-0.96$) and is moderately anti-correlated in the arms ($r=-0.47$), consistent
with strong internal Faraday depolarization in dense, high-emission-measure regions and increased magnetic disorder in star-forming environments.

The comparison between the two wavelength regimes shows that
radio polarized intensity grows with increasing $N_{\rm H}$, but
not FIR polarized intensity.
At 154\,\micron, the
lacking
dependence reflects the increasing dust column and the stronger radiation fields required to maintain grain alignment in dense environments.
At 6\,cm, the correlation primarily traces enhanced magnetic field strength and cosmic-ray electron densities.
Yet in both tracers, the fractional polarization drops with $N_{\rm H}$ because high-density regions host more tangled fields, more turbulence, and stronger depolarization
(either line-of-sight for radio or structural/turbulent for FIR).
This shared decline in fractional polarization despite the different emission mechanisms highlights the dominant role of magnetic field disorder and environmental complexity
in dense gas.

The trends we find for NGC\,6946 strongly echo those reported for M51 in the SOFIA Legacy Program \citep{SALSAI}.
In M51, FIR polarized intensity at 154\,\micron\
does not
increase with gas surface density while FIR fractional polarization decreases sharply with increasing $N_{\rm H}$ and star-formation indicators.
Likewise, the polarized radio synchrotron emission in M51 rises with gas density, but the fractional polarization decreases in dense, actively star-forming regions due to
Faraday depolarization and field disorder.
\cite{SALSAI}~interpret this as evidence that dust polarization traces magnetic fields within higher-density molecular gas layers where turbulence and cloud-scale disorder
grow with gas column while the radio polarization traces the more diffuse magnetized disk.
Our results for NGC\,6946 match this picture closely:
FIR polarization becomes increasingly suppressed in high-column-density environments, whereas radio polarized intensity continues to track the strengthening field without
preserving a high polarization fraction.

The central region and spiral arm behaviors in NGC\,6946 also mirror the radial and environmental trends in M51.
In both galaxies, dense central regions show a higher rise in total and polarized intensity with gas density but suffer the strongest suppression in fractional polarization.
The arm regions show moderate
or no
correlation in these tracers, reflecting the intermediate turbulence and magnetic-ordering conditions of spiral arms.
Thus, our analysis reinforces the broader SOFIA result that FIR and radio polarization respond differently to gas density but converge in showing weaker polarization
fractions where the ISM becomes dense, turbulent, and magnetically complex.

\subsection{Origin of the FIR dust polarization}\label{sec:fir_origin}

Within the RAT framework introduced in Sect.~\ref{sec:intro}, the efficiency of grain alignment, and hence the observed FIR polarization fraction, depends on
grain size, the intensity and anisotropy of the radiation field, gas density (which damps grain rotation via collisions), and the local magnetic field strength
\citep{Vaillancourt2009}. Stronger, more anisotropic radiation fields increase grain spin-up and enhance alignment, although very intense fields can instead disrupt or
heat grains, reducing alignment efficiency \citep{Hoang2016}. Higher gas density increases collisional damping and is often associated with more tangled fields or
multiple layers along the line of sight, both of which suppress the observed polarization fraction even where dust mass, and hence FIR intensity, is high. Consistent
with this, the SALSA program has found that FIR polarization often reveals more complex, multi-mode field structures than radio synchrotron observations
\citep{Surgent2023}.

\subsection{Polarization fraction versus dust temperature at 154\,\micron}

\begin{figure}
    \centering

   \includegraphics[width=0.98\columnwidth]{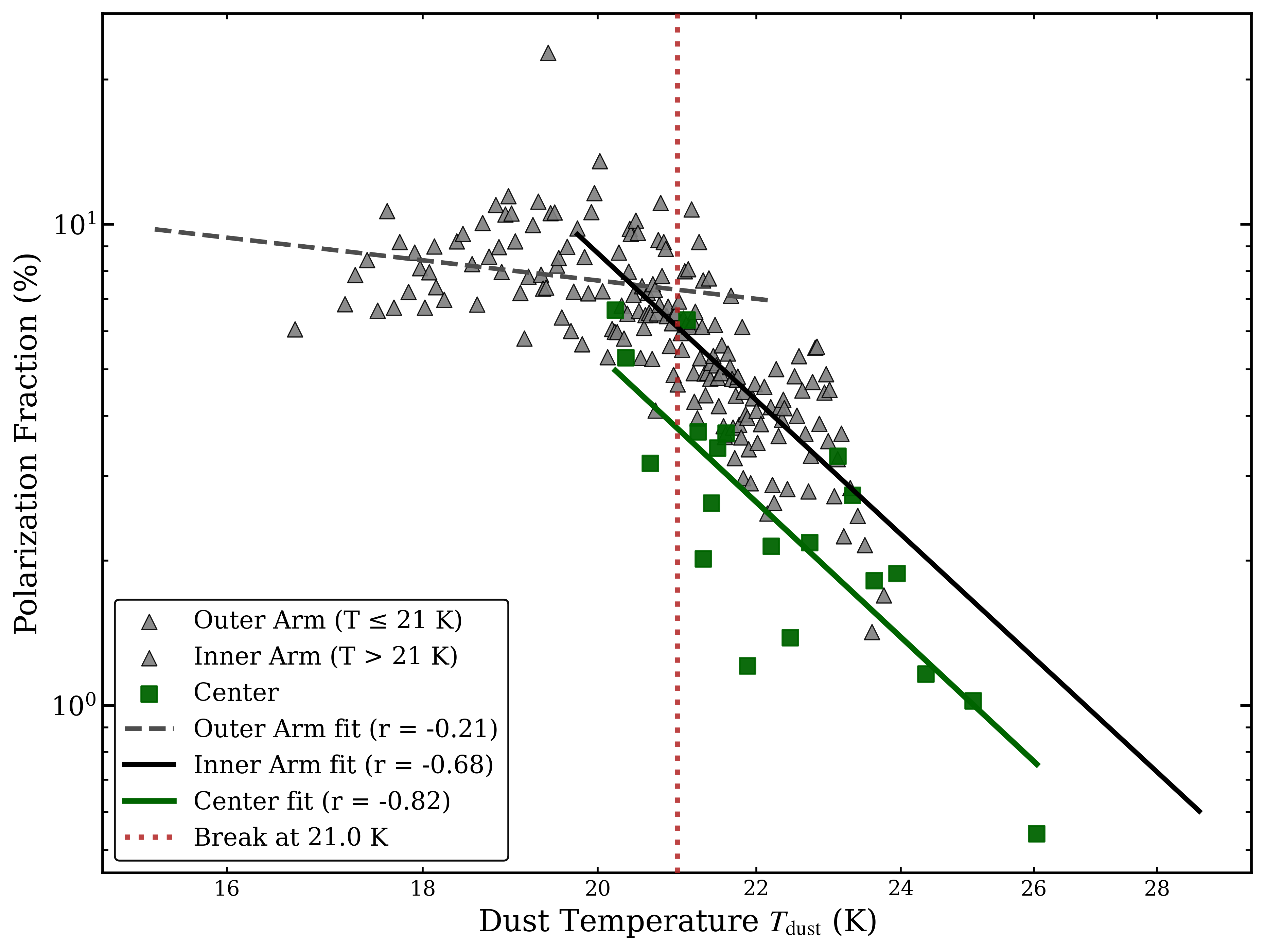}
    \caption{Polarization fraction $p_{dust}$ at 154\,\micron\ as a function of dust temperature $T_{\rm dust}$ in NGC\,6946.
    Green filled squares denote the central region (within $\sim 1$\,kpc), filled grey triangles the inner spiral arms ($T_{\rm dust} > 21$\,K), and open grey triangles the
    outer spiral arms ($T_{\rm dust} \leq 21$\,K).
Dashed grey, solid black, and solid green lines show power-law fits to the outer-arm, inner-arm, and central region populations, respectively. The vertical red dashed line
marks the clear break at $T_{\rm dust} = 21.0$\,K separating the regimes of efficient and thermally disrupted grain alignment via radiative torques \citep{Hoang2019}.
Above this temperature, the steep decline of polarization traces the rapid loss of dust alignment in the warmer, star formation-dominated inner galaxy, whereas the flatter
relation in the colder outer arms reflects sustained high alignment efficiency.}
\label{fig:pfvsT_dust}
\end{figure}

Figure~\ref{fig:pfvsT_dust} shows the polarization fraction $p_{dust}$ at 154\,\micron\ as a function of dust temperature $T_\mathrm{dust}$ derived from $U_{min}$ map.
Three distinct populations are evident: the central region (green squares), the inner spiral arms ($T_\mathrm{dust} > 21$\,K; filled grey triangles), and the outer spiral
arms ($T_\mathrm{dust} \leq 21$\,K; open grey triangles).

The central and inner-arm regions follow a tight anti-correlation, whereas the outer arms are significantly flatter.
A clear break occurs at $T_\mathrm{dust} \approx 21$\,K (vertical red dashed line), above which the polarization fraction drops much more rapidly with increasing temperature.

This behavior is a direct manifestation of the temperature-dependent alignment efficiency of non-spherical dust grains with the interstellar magnetic field.
The fractional polarization produced by thermally rotating, magnetically aligned grains can be expressed as
\begin{equation}
p \approx p_0 \, f_\mathrm{align} \, C_\mathrm{pol},
\end{equation}
where $p_0$ is the intrinsic polarization efficiency of a perfectly aligned grain (typically 0.1--0.2 for silicate grains at 154\,\micron), $f_\mathrm{align}$ is the fraction
of grains with suprathermal rotation aligned with $\mathbf{B}$, and $C_\mathrm{pol} \lesssim 1$ is a geometric depolarization factor arising from variations of the magnetic
field direction along the line of sight \citep{Andersson2015, Hoang2019}.

Radiative torque (RAT) theory predicts that grain alignment is efficient when the radiative precession timescale is shorter than the gas-damping timescale, which translates
to a critical temperature above which alignment is lost because thermal fluctuations within the grain become comparable to the magnetic alignment energy
\citep{hoang2019grain}.
For typical ISM conditions in spiral galaxies, the drop in alignment efficiency becomes pronounced around $T_\mathrm{dust} \sim 20$--$25$\,K, exactly the range we observe in
NGC\,6946.

In the warmer inner arms and central region ($T_\mathrm{dust} > 21$\,K), the rapid decline of $p$ with increasing $T_\mathrm{dust}$ therefore reflects the progressive loss of
grain alignment due to enhanced internal thermal (phonon) fluctuations and randomization of grain angular momentum (``RAT disruption''; \citealt{lazarian2007radiative}).
Conversely, the outer arms remain colder and are dominated by molecular clouds and diffuse H\,\textsc{i} regions illuminated by a weaker radiation field; in these
environments grains remain efficiently aligned even up to dust temperatures of $T_{\mathrm{dust}} \sim 21\,\mathrm{K}$, producing a much flatter $p$--$T_{\mathrm{dust}}$
relation. The steeper slope in the central region compared to the inner arms ($-0.81$ versus $-0.68$) may further indicate additional depolarization from tangled magnetic
fields in the central starburst region, where turbulence injected by supernovae and stellar feedback is strongest \citep{SALSAII}. Thus, the observed break at
$T_{\mathrm{dust}} \approx 21\,\mathrm{K}$ simultaneously probes both the microphysics of dust alignment and the macroscopic variation of star-formation intensity across the
galaxy.

\subsection{Polarization fraction versus CO gas density }

\begin{figure}
    \centering
    \includegraphics[width=0.45\textwidth]{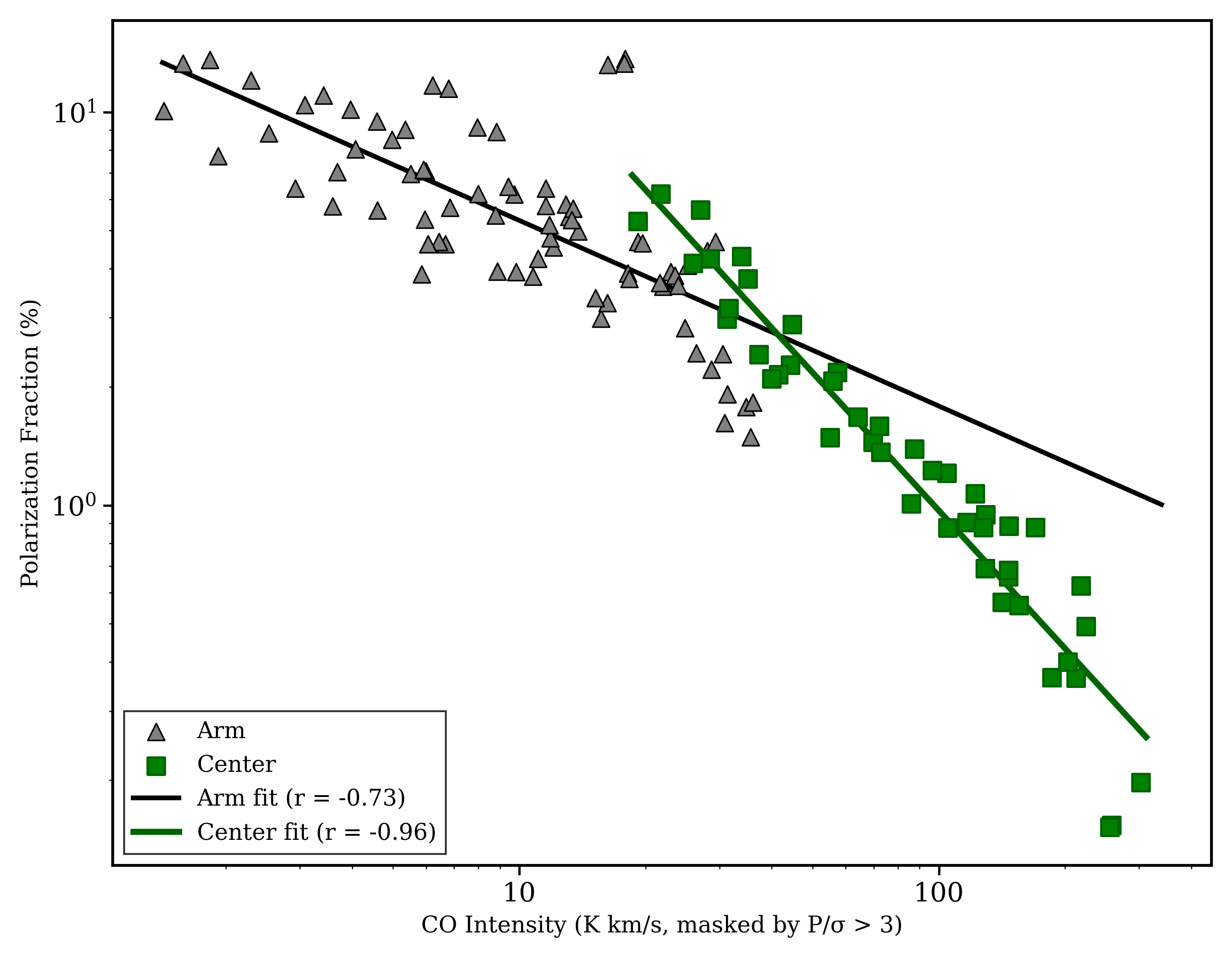}
    \caption{Polarization fraction ($p_{dust}$) versus CO intensity for the arm (grey) and central (green) regions of NGC\,6946. A clear anti-correlation is observed, with
    $r_p = -0.73$ (arm) and $r_p = -0.96$ (central region), indicating that stronger molecular emission corresponds to lower polarization. The trend reflects reduced grain
    alignment efficiency in denser regions due to enhanced collisional damping, radiation field attenuation, and magnetic field tangling, with the steeper decline in the
    central region, suggesting stronger depolarization and grain disruption effects.}
    \label{fig:pfvsco}
\end{figure}

The relationship between the 154\,$\mu$m polarization fraction $p_{dust}$ and the CO intensity shown in Figure~\ref{fig:pfvsco} reveals a strong anti-correlation in both the
spiral arms ($r_p=-0.73$) and, even more strikingly, in the central region ($r_p=-0.96$), indicating that polarization efficiency declines rapidly with increasing molecular
gas column density. This behavior is naturally explained within the framework of Radiative Alignment Torque (RAT) theory \citep{Dolginov1976, Lazarian2007}, which requires
dust grains to reach suprathermal rotation in order to align with the magnetic field. The condition for alignment is that the radiative-torque angular velocity exceeds the
thermal angular velocity, $\omega_T=\sqrt{k_{\rm B}T_{\rm gas}/I(a)}$, such that $\omega_{\rm RAT}/\omega_T \gtrsim 1$ \citep{Lazarian2007}. Equating these velocities defines
the minimum aligned grain size $a_{\rm align}$ through
\begin{equation}
    \omega_{\rm RAT}\!\left(a_{\rm align},U_{\rm rad}\right)
    \;=\;
    \omega_{T}\!\left(a_{\rm align},n_{\rm H},T_{\rm gas}\right),
\end{equation}
leading to the scaling
\begin{equation}
    a_{\rm align}
    \propto
    \mathcal{F}\!\left(
        U_{\rm rad},\, n_{\rm H},\, T_{\rm gas},\,
        \rho_{\rm grain},\, Q_{\rm RAT}
    \right),
\end{equation}
where high gas densities ($n_{\rm H}$) enhance gas damping and increase $a_{\rm align}$, while stronger and more anisotropic radiation fields ($U_{\rm rad}$) reduce it
\citep{Hoang2019}. In the spiral arms, the radiation field remains sufficiently strong and anisotropic that $a_{\rm align}$ stays small, allowing a substantial fraction of
grains to remain aligned despite increasing CO intensity consistent with the moderately steep arm correlation. In contrast, the central region contains much higher molecular
gas densities, stronger turbulence, and increased magnetic field disorder \citep{Beck2007, Tabatabaei+2013}, all of which raise $a_{\rm align}$ and greatly reduce the
population of aligned grains, producing the nearly monotonic decline of $p$ with CO ($r_p=-0.96$). Furthermore, the intense nuclear radiation field may drive grains into the
Radiative Torque Disruption (RAT--D) regime \citep{Hoang2019}, in which centrifugal stresses fragment the largest, most efficiently aligned grains, further suppressing the
polarization fraction. Therefore, the steeper central trend reflects the combined effects of increased alignment-size thresholds, magnetic depolarization, and possible RAT--D
grain destruction in the dense, highly irradiated nucleus of NGC\,6946.

\section{Conclusions}
\label{sec:conclusions}

In this work, we presented a comparative analysis of far-infrared (FIR) and radio polarization in NGC\,6946, aiming to investigate the behavior of magnetic fields across
different phases of the interstellar medium (ISM). By combining SOFIA/HAWC+ observations at 154~$\mu$m with radio continuum data at 6~cm, we probed magnetic fields in the
cold dust-emitting and the warm diffuse synchrotron-emitting components of the galaxy.

Our main results can be summarized as follows:

\begin{enumerate}

\item The synchrotron polarization reveals a large-scale magnetic field structure, with higher polarization fractions in interarm regions and significant
    depolarization in spiral arms. In contrast, the FIR polarization shows lower overall polarization fractions and stronger spatial variations, indicating a more complex
    magnetic field structure in the cold ISM.

\item Both FIR and synchrotron polarization fractions decrease with increasing star formation activity, suggesting that turbulence driven by star formation reduces the
    degree of magnetic field ordering. This trend is consistent with the expected impact of stellar feedback on the ISM.

\item A comparison between dust and synchrotron polarization shows no correlation in polarized intensity, while the fractional polarizations exhibit a
    weak-to-moderate correlation in the spiral arms and a strong correlation in the central region. This indicates that, although the emission mechanisms differ, the degree
    of magnetic field ordering is coupled across ISM phases.

\item Radial trends show that the synchrotron polarization fraction is strongly suppressed in the central kiloparsec, holds at $\sim 13$--$17\%$ across the mid disk, and tends to increase to $\sim 25$--$30\%$ beyond $R \approx 7$\,kpc, reflecting the growing dominance of the ordered interarm field. The FIR polarization fraction can be measured reliably only out to $R \approx 5.4$\,kpc; over this range it remains low and approximately constant at $\sim 2$--$3\%$, indicating that magnetic field ordering in the cold dusty ISM is modulated by local environmental conditions rather than by a global radial gradient.

\end{enumerate}

Overall, our results indicate that NGC\,6946 hosts a globally ordered magnetic field that is locally modulated by turbulence. The
combination of FIR and radio polarimetry provides a powerful tool to disentangle magnetic field properties in different ISM phases and offers new constraints on the coupling
between magnetic fields and galaxy evolution processes.

Future high-resolution polarimetric observations with next-generation facilities will further improve our ability to probe magnetic field structure across multiple scales and
environments, enabling a more comprehensive understanding of the role of magnetic fields in shaping the ISM of galaxies.

\section*{Acknowledgement}
\begin{acknowledgments}
Based on observations made with the NASA/DLR Stratospheric Observatory for Infrared Astronomy (SOFIA). SOFIA is jointly operated by the Universities Space Research Association, Inc. (USRA), under NASA contract NNA17BF53C, and the Deutsches SOFIA Institut (DSI) under DLR contract 50\,OK\,2002 to the University of Stuttgart.

E.L.-R.\ thanks the support of the NASA Astrophysics Decadal Survey Precursor Science (ADSPS) Program (NNH22ZDA001N-ADSPS) with ID 22-ADSPS22-0009 and agreement number 80NSSC23K1585.

F.S.T.\ acknowledges the support from the DYNAVERSE Cluster of Excellence (Cologne, Bonn), which is funded by the Deutsche Forschungsgemeinschaft (DFG, German Research Foundation) under Germany's Excellence Strategy, EXC 3037-533607693.
\end{acknowledgments}

\bibliography{references}{}
\bibliographystyle{aasjournal}

\appendix

\section{Modeling Dust Heating in NGC\,6946: Understanding the $U_{\min}$ Map}\label{app:umin}
To obtain the radiation field across NGC\,6946, \cite{Aniano2012} adopted the \citet{Draine2007} dust model. In this framework, most of the dust mass is heated by a minimum
radiation field $U_{\min}$, while a smaller fraction is exposed to stronger fields up to $U_{\max}$. The dust mass distribution over starlight intensity is expressed as:
\begin{equation}
\label{eq:DL07}
\begin{aligned}
\frac{dM_{\rm dust}}{dU}
= M_{\rm dust}\Big[(1 - \gamma)\,\delta(U - U_{\min}) \\
\qquad\qquad +\, \gamma\,\frac{\alpha - 1}{U_{\min}^{\,1-\alpha} - U_{\max}^{\,1-\alpha}}\,U^{-\alpha}\Big],
\end{aligned}
\end{equation}
where $\gamma$ is the fraction of dust heated by the power-law component and $\alpha$ is its slope. This two-component heating model enables the warm tail of the infrared SED
to be reproduced while most of the dust mass is kept near equilibrium with the diffuse radiation field.
\cite{Aniano2012} fitted the \citet{Draine2007} model to each pixel of the observed multi-band IR and submillimeter maps, which were convolved to a common resolution of
18$\arcsec$ to determine $U_{\min}$, $\gamma$, $q_{\rm PAH}$, and the dust mass surface density. Among these parameters, $U_{\min}$ is particularly useful because it captures
the intensity of the ambient radiation field heating the bulk of the dust. While its absolute value can be affected by calibration and modeling uncertainties
\citep{Aniano2012, Aniano2020}, its spatial variations reliably trace real changes in the interstellar radiation environment.

\section{Separating Thermal and Nonthermal Components at
\texorpdfstring{6\,cm}{6 cm}}\label{TNT}

The observed radio continuum emission at 6\,cm is a combination of two components: the thermal free-free emission and the nonthermal synchrotron radiation. Because only the
nonthermal component is polarized, mapping the polarization fraction in radio will be erroneous if the thermal contamination is not subtracted.

Following \cite{Tabatabaei+2013}, we separated the thermal and nonthermal components of the 6\,cm emission using a thermal radio tracer.
As the brightest hydrogen recombination line, H$\alpha$ emission is linearly proportional to the free-free emission in a thermally ionized medium and, hence, provides a
physically motivated tracer for the thermal radio continuum. However, the H$\alpha$ emission should be corrected for dust extinction. Taking advantage of the dust mass
distribution derived from a detailed dust SED analysis in NGC\,6946, \cite{Tabatabaei+2013} mapped the optical depth of dust at the H$\alpha$ wavelength and obtained a
de-reddened H$\alpha$ map at 18\as as a thermal template. We used this map to obtain the thermal free-free emission at 6\,cm (4.85\,GHz).

The H$\alpha$ emitting medium in galaxies is often optically thick to ionizing Lyman photons, and hence, the case-B recombination applies \citep{oster}. Under this
condition, and taking into account that the ISM in NGC\,6946 is optically thin to the free-free radiation at 6\,cm (4.85\,GHz), the following relation holds between the
free-free brightness temperature ($T_b$) in Kelvin and H$\alpha$ intensity ($I_{\rm H\alpha}$) in erg\,s$^{-1}$\,cm$^{-2}$\,sr$^{-1}$.
\begin{equation}
    T_b = 3.484\,\times 10^4\,a\,\nu_{\rm GHz}^{-2.1}\,T_{e4}^{0.667}\,10^{\frac{0.029}{T_{e4}}}\, (1\,+\,0.08)\, I_{\rm H\alpha},
\end{equation}
where $T_{e4}$ is the electron temperature in units of $10^4$\,K and $a\simeq1$ at cm wavelengths. The factor (1 + 0.08) takes
into account the contribution from singly ionized He. We obtained the intensity of
the free-free emission in mJy/beam at 6\,cm assuming $T_e=10^4$\,K. As discussed by \cite{Tabatabaei+2013}, smaller values of $T_e$ are reported from measurements in the
Milky Way (e.g., Haffner et al. 1999; Madsen et al. 2006); a
variation of $T_e$ by 2000--3000\,K would change the thermal fraction by less than $\simeq$20\%. We also note that a fixed electron temperature is supported by the shallow
metallicity gradient found in NGC\,6946 \citep{mous}. Subtracting
the free-free emission from the observed 6\,cm results in a map of the synchrotron emission, $I_{\rm syn}$.

\begin{figure*}
    \centering
    \includegraphics[width=0.45\textwidth]{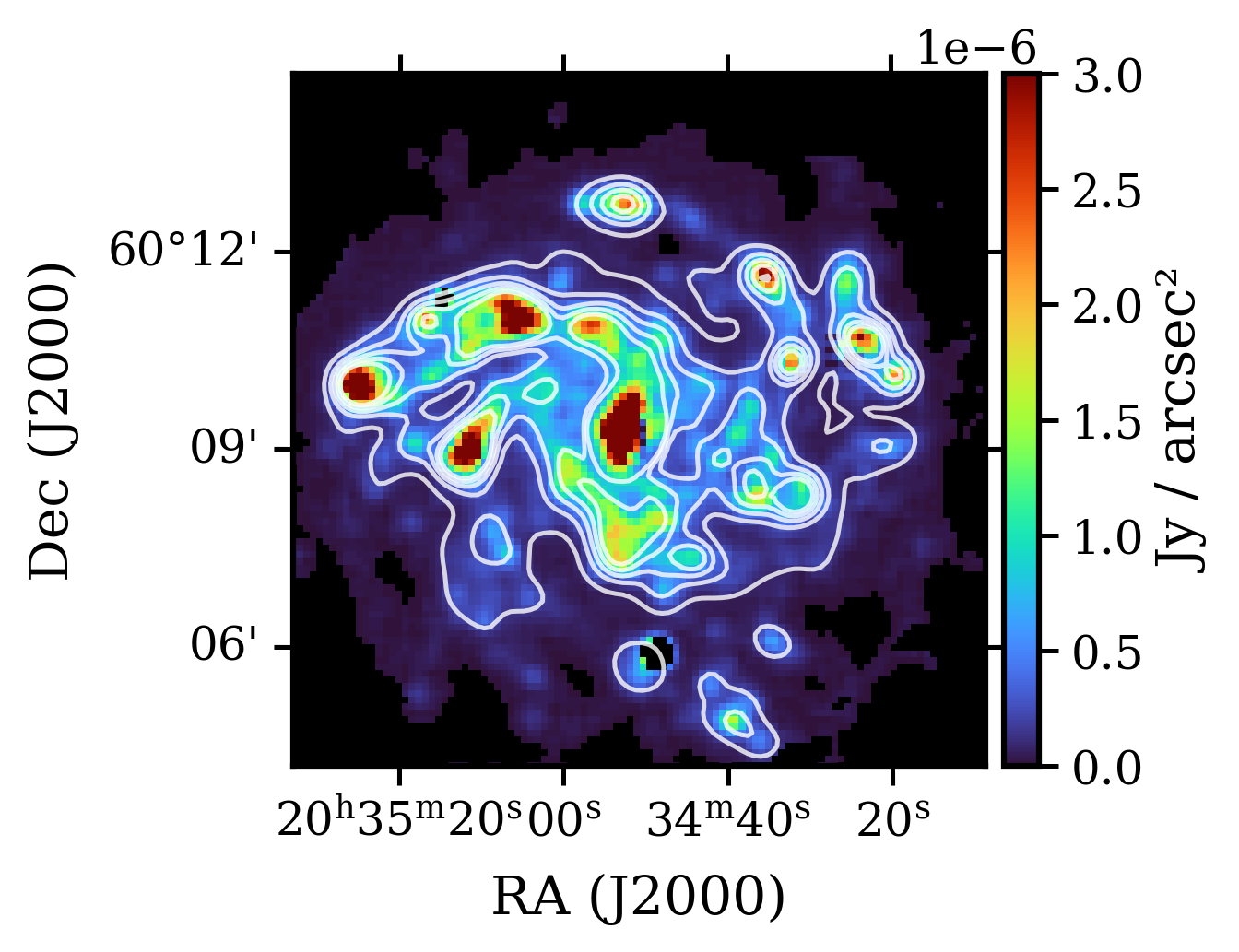}
    \includegraphics[width=0.45\textwidth]{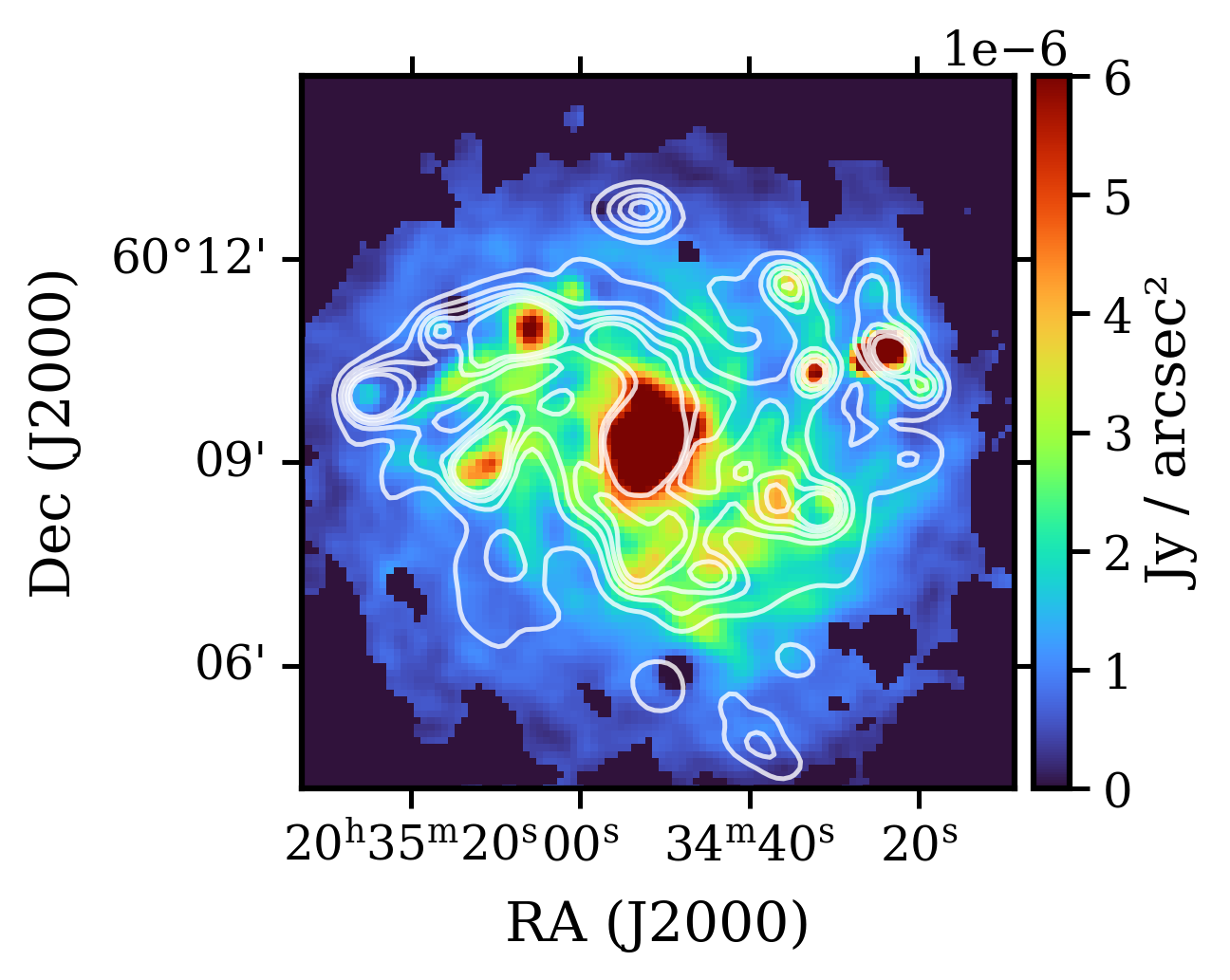}
    \caption{
    Left: Thermal radio emission at 6\,cm (color), with contour levels of 0.05, 0.10, 0.20, 0.30, 0.40, 0.50 $\times$ $\Sigma_{SFR}$.
    Right: Total synchrotron radio emission at 6\,cm (color), with contour levels of 0.05, 0.10, 0.20, 0.30, 0.40, 0.50 $\times$ $\Sigma_{SFR}$.
    }
    \label{fig:thermal_non_thermal}
\end{figure*}

Figure \ref{fig:thermal_non_thermal} shows the thermal and nonthermal maps of NGC\,6946 at 6\,cm. Both the thermal and nonthermal emissions are brightest in the center and
star-forming regions. Unlike the thermal emission, significant extended emission is found in the nonthermal map, which is expected due to the synchrotron radiation of CREs
diffused over the disk and in the halo. Figure~\ref{fig:thermal_non_thermal} presents the thermal (free--free) radio continuum emission of NGC\,6946 at 6\,cm and $18''$
resolution, shown in color and overlaid with contours of the star formation rate (SFR). The thermal emission was obtained by subtracting the nonthermal synchrotron
component from the total intensity map after applying a $3\sigma$ threshold to both datasets, using rms noise levels of $22\,\mu$Jy\,beam$^{-1}$ for the total intensity and
$35\,\mu$Jy\,beam$^{-1}$ for the synchrotron emission. This masking ensures a robust separation of the thermal component and suppresses spurious values in low signal-to-noise
regions. The thermal emission closely follows the SFR distribution, with the highest intensities located along the spiral arms and in the central region. The peak thermal
surface brightness reaches $\sim 3\times10^{-6}$~Jy\,arcsec$^{-2}$ in prominent star-forming complexes, while the emission decreases to $\lesssim
3\times10^{-7}$~Jy\,arcsec$^{-2}$ toward the outer disk. The SFR contours trace the same structures seen in the thermal map, confirming that the free--free radio emission is
dominated by ionized gas associated with ongoing massive star formation.

\begin{figure*}
    \centering
    \includegraphics[width=0.7\textwidth]{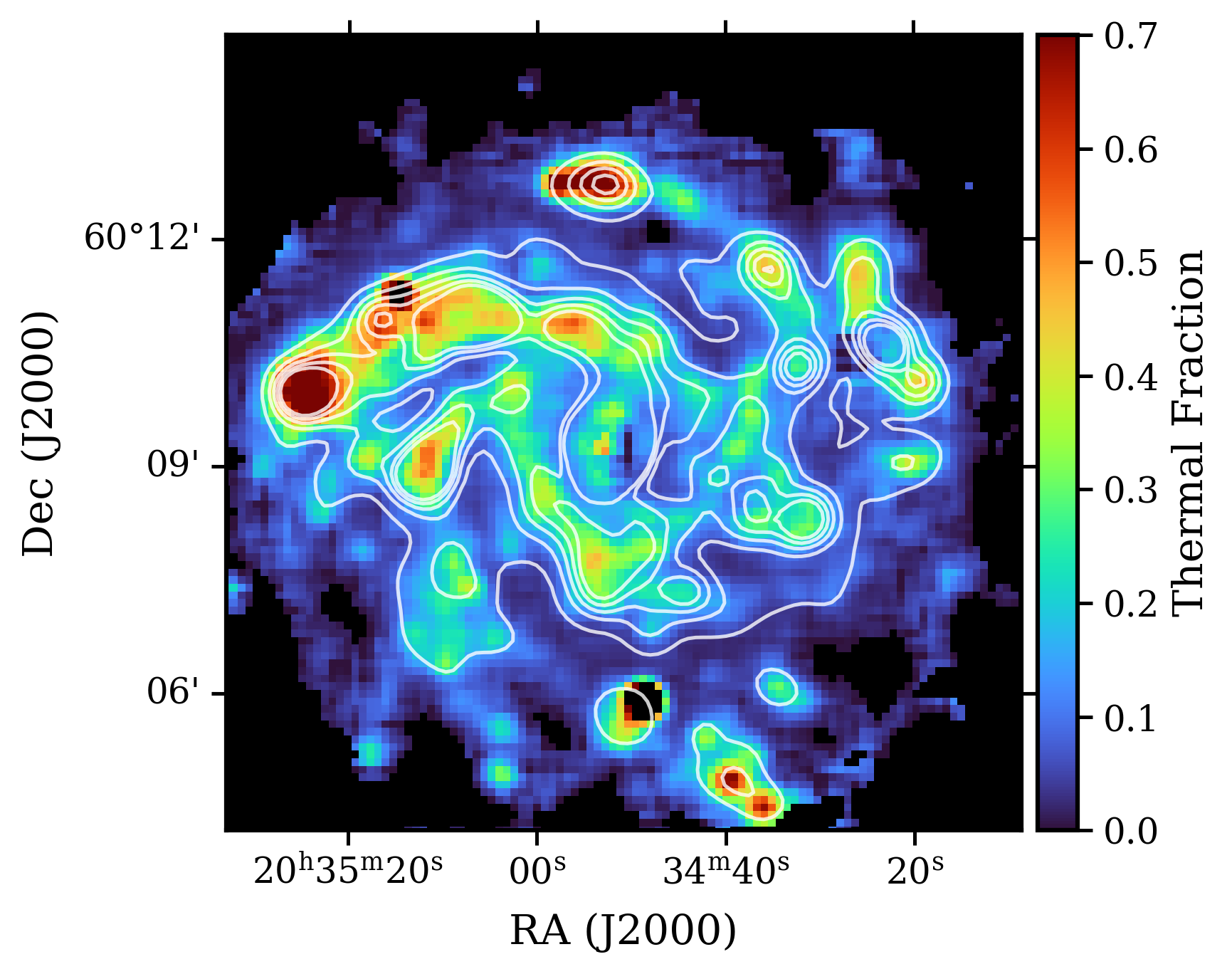}
    \caption{Thermal fraction at 6\,cm (color).
    Contours indicate the surface-density levels of star formation rate (SFR), with contour levels of 0.05, 0.10, 0.20, 0.30, 0.40, 0.50 $\times$ $\Sigma_{SFR}$.}
    \label{fig:thermal_frac}
\end{figure*}

Figure~\ref{fig:thermal_frac} shows the thermal fraction at 6\,cm resolution ($18''$) in NGC\,6946, displayed in color and overlaid with contours of the star formation rate
(SFR). The thermal fraction was computed as the ratio of the thermal (free--free) emission to the total radio continuum intensity after masking both maps below $3\sigma$,
adopting rms noise levels of $22\,\mu$Jy\,beam$^{-1}$ for the total intensity and $35\,\mu$Jy\,beam$^{-1}$ for the synchrotron component. This ensures that only regions with
reliable signal contribute to the fraction map. The thermal fraction exhibits strong spatial variations across the disk, with the highest values ($f_{\rm th}\gtrsim0.6$)
found in prominent star-forming regions along the spiral arms and in the central area, coincident with peaks in the SFR contours. Lower thermal fractions ($f_{\rm
th}\lesssim0.1$) dominate the interarm regions and the outer disk, where synchrotron emission becomes increasingly important. The close correspondence between enhanced
thermal fraction and elevated SFR confirms that the free--free contribution is closely linked to ongoing massive star formation.

\begin{table}[htbp]
\centering
\caption{Regional thermal (free--free) flux densities and thermal fractions in NGC\,6946. These are averages over the full arm and central region masks, and are therefore
lower than the local peak values ($f_{\rm th}\gtrsim0.6$) reached in the brightest individual star-forming complexes quoted above.}
\begin{tabular}{lcc}
\hline
Region & Flux density (mJy) & Thermal fraction ($f_{\mathrm{th}}$) \\
\hline
Arm region & 67.23 $\pm$ 0.20 & 0.216 $\pm$ 0.003 \\
Central region & 20.90 $\pm$ 0.51 & 0.159 $\pm$ 0.006 \\
\hline
\end{tabular}
\end{table}

\section{Multi-wavelength Maps of NGC\,6946}

This appendix presents a set of multi-wavelength maps of NGC\,6946 that provide additional context for the analysis discussed in the main text. These maps illustrate key
tracers of star formation, gas content, magnetic field orientation, and interstellar radiation field properties across the galaxy.

\begin{figure*}
\centering
\includegraphics[width=\textwidth]{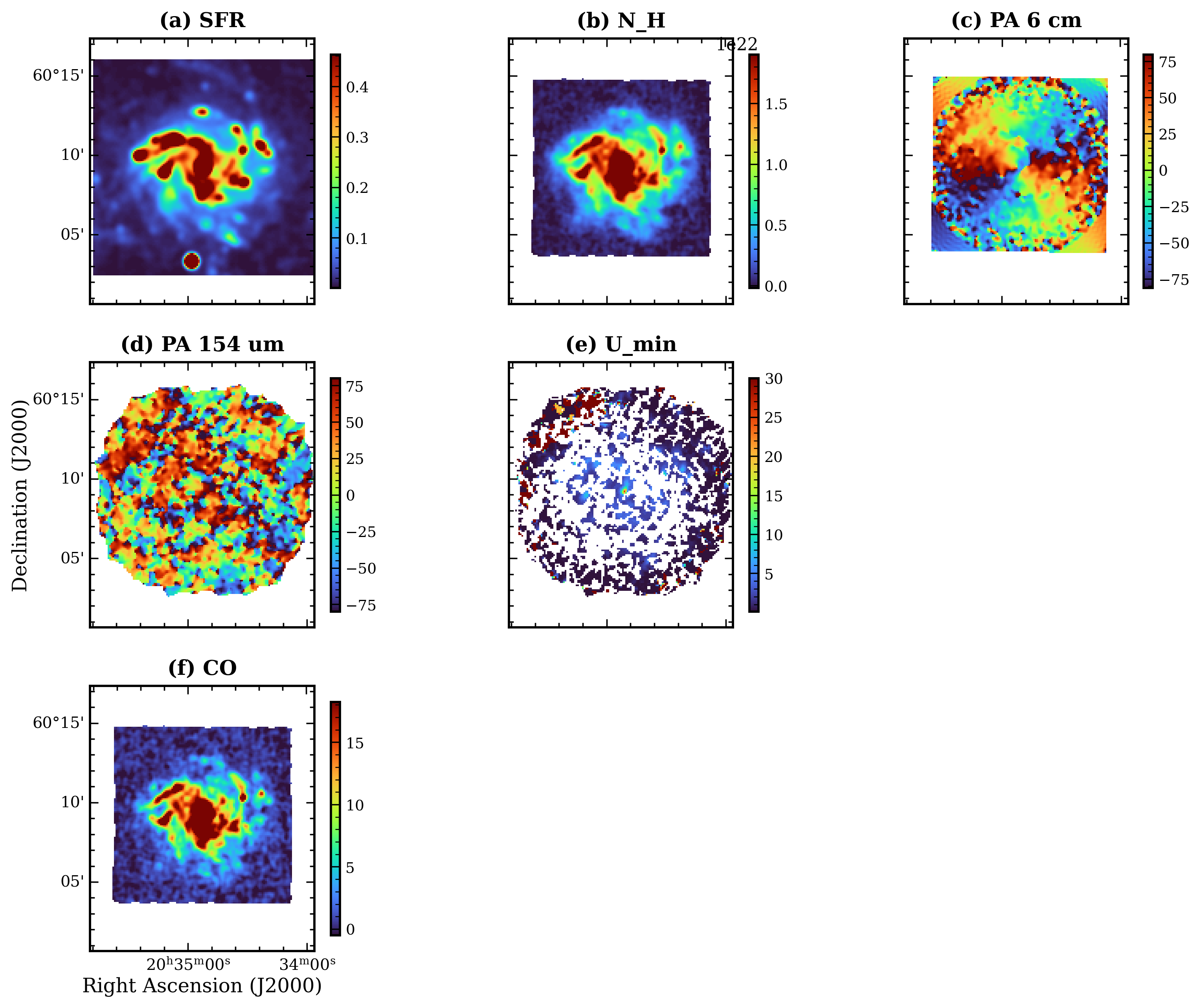}
\caption{
Multi-wavelength maps of NGC\,6946. Panels show (a) SFR surface density, (b) total hydrogen column density, (c) magnetic field position angle at 6~cm, (d) magnetic field
position angle at 154~$\mu$m, (e) minimum radiation field $U_{\min}$, and (f) integrated CO intensity. All maps are shown at a common resolution of $18^{\prime\prime}$ and in
the same coordinate system. Only regions above the adopted signal-to-noise thresholds are displayed.
}
\label{fig:allmaps}
\end{figure*}

Figure~\ref{fig:allmaps} shows the following quantities:
(a) the star formation rate (SFR) surface density map, highlighting regions of active star formation;
(b) the total hydrogen column density ($N_{\mathrm{H}}$), tracing the distribution of neutral gas;
(c) the position angle (PA) of the magnetic field derived from 6~cm radio polarization data;
(d) the magnetic field position angle derived from 154~$\mu$m far-infrared (FIR) polarization;
(e) the minimum radiation field intensity ($U_{\min}$), which characterizes the diffuse interstellar radiation field heating the bulk of the dust; and
(f) the integrated CO intensity map, tracing the molecular gas component.

All maps are presented in equatorial coordinates (J2000) and have been convolved to a common angular resolution of $18^{\prime\prime}$ to enable a consistent comparison
across different tracers. The data are drawn from a combination of facilities, including SOFIA/HAWC+ for FIR observations, VLA and Effelsberg for radio data, and ancillary
surveys such as HERACLES (CO) and THINGS (H\,{\sc i}).

These maps provide a comprehensive overview of the multi-phase ISM in NGC\,6946 and illustrate the spatial relationships between star formation activity, gas distribution,
radiation field intensity, and magnetic field structure. In particular, they support the interpretation of the correlations explored in Sect.~\ref{sec:discussion} by visually linking the
different ISM components and highlighting the regions used in the analysis (e.g., spiral arms and central region).

\end{document}